\documentclass[useAMS,usenatbib]{mn2e}
\usepackage{graphicx}
\usepackage{subfigure}
\usepackage[english]{babel}
\usepackage{amssymb}

\def\hi{ \ifmmode {\mbox H{\scshape i}}\else H{\scshape i} \fi}
\def\h2{\ifmmode {\mbox H$_2$}\else H$_2$\fi}
\def\cosmos{\ifmmode {\mbox {\it COSMOS}}\else {\it COSMOS} \fi}

\title[Gas evolution in galaxies at $0.5\leq z \leq 2.0$]{An indirect
    measurement of gas evolution in galaxies at $0.5\leq z \leq 2.0$}
\author[Popping et al.]{G. Popping$^{1}$\thanks{E-mail:
    G.Popping@astro.rug.nl}, K.I. Caputi$^{2,1}$, R.S. Somerville$^{3}$, and  S.C. Trager$^{1}$\\
$^{1}$Kapteyn Astronomical Institute, University of Groningen, Postbus 800, NL-9700 AV Groningen, the Netherlands\\
$^{2}$Scottish Universities Physics Alliance (SUPA), Institute for Astronomy, The University of Edinburgh,\\
Royal Observatory, Edinburgh EH9 3HJ\\
$^{3}$Department of Physics and Astronomy, Rutgers University, 136
Frelinghuysen Road, Piscataway, NJ 08854, USA}
\begin{document}

\maketitle

\begin{abstract}
One key piece of information missing from high redshift galaxy surveys
is the galaxies' cold gas contents. We present a new method to
indirectly determine cold gas surface densities and integrated gas
masses from galaxy star formation rates and to separate the atomic and
molecular gas components. Our predicted molecular and total gas
surface densities and integrated masses are in very good agreement
with direct measurements quoted in the literature for low and high-z
galaxies. We apply this method to predict the gas content for a sample
of $\sim 57000$ galaxies in the \cosmos field at $0.5<z<2.0$, selected
to have $I_{AB} < 24$ mag. This approach allows us to investigate in
detail the redshift evolution of galaxy cold and molecular gas content
versus stellar mass and to provide fitting formulae for galaxy gas
fractions. We find a clear trend between galaxy gas fraction,
  molecular gas fraction and stellar mass with redshift, suggesting
  that massive galaxies consume and/or expel their gas at higher
  redshift than less massive objects and have lower fractions of their
  gas in molecular form. The characteristic stellar mass separating
  gas- from stellar-dominated galaxies decreases with time. This
  indicates that massive galaxies reach a gas-poor state earlier than
  less massive objects. These trends can be considered to be another
  manifestation of downsizing in star formation activity.
\end{abstract}


\section{Introduction}
In a pioneering work, \citet{Madau1996} demonstrated that the star
formation (SF) activity of the Universe peaks at a redshift $z\sim1-3$
and decreases at lower redshifts to its present day value. In the
following years, many studies confirmed the existence of a peak in the
star formation rate (SFR) density around these redshifts, and a clear
decline at $z<1$ \citep[e.g.,][]{Hopkins2004,Hopkins2006}. This makes
the epoch between $z\sim2$ and $z\sim0.5$, in which massive galaxies stop
forming the bulk of their stars and become passive, a crucial period
for galaxy evolution.

One of the features which has become clear in this period of galaxy
evolution is that the most massive objects (giant elliptical galaxies
hosted in galaxy groups and clusters) formed their stars early, while
less massive objects continued to form stars until the present (this
trend is commonly called `galaxy downsizing'). This behavior has
many different observational manifestations
\citep[e.g.,][]{Faber1992,Worthey1992,Cowie1996,Trager2000,Drory2004,Drory2005,Cimatti2006,Trager2008},
but its physical origin remains unclear \citep{fontanot2009}.

The star formation rate in galaxies is closely linked to the galaxy
gas content. Observations have shown that SF in the Milky Way takes
place in dense, massive and cold giant molecular clouds
\citep[GMC;][]{Solomon1987,Bolatto2008,McKee2007}. This makes the SF
tightly connected to the molecular and atomic gas
available. \citet{schmidt1959} found a power-law relation between the
surface density of SFR and gas surface density. This work has become
the cornerstone of a wealth of studies relating SFR surface densities
and cold gas and molecular hydrogen surface densities
\citep[e.g.,][]{Kennicutt1998law,Bigiel2008,Schruba2011}. Both the
observations of GMCs, and the power-law relations between SFR and gas
densities demonstrate that information about the gas and its partition
into atomic and molecular hydrogen is essential for a proper
understanding of galaxy evolution and the build up of stellar
mass. 

An essential landmark and significant step forward in the study of
galaxy evolution at higher redshifts has been the development of large
samples of distant galaxies with extensive multi-wavelength
information (e.g., {\it GOODS} and \cosmos). These samples comprise
imaging data from X-rays to the infrared, which allow us to derive
numerous physical properties of galaxies, such as stellar masses and
star formation rates. However, one of the key pieces of information
missing in these studies is the cold gas content. Deriving the amount
of cold gas present as a function of stellar mass for large
representative samples would allow us to better understand the history
of gas consumption and the overall phenomenon of galaxy downsizing.

Information about the gas content of galaxies
  also provides important constraints on theoretical models of galaxy
  formation. Gas fractions help to break the
  degeneracies in different physical mechanisms that are included in
  these models, namely star formation and stellar feedback (Caviglia
  \& Somerville, in prep).
Moreover, cosmological galaxy formation models have begun to include recipes
that allow a detailed tracking of atomic and molecular gas with
redshift in the context of general galaxy properties \citep[e.g., gas
  mass functions, gas properties as a function of stellar
  mass;][Popping et al., in prep]{Obreschkow2009,Fu2010,Gnedin2011,Krumholz2011,Kuhlen2011,Lagos2011sflaw,Lagos2011cosmic_evol}. 
Observational constraints on atomic and molecular gas content as a
function of galaxy properties and cosmic time are crucial for
developing and calibrating these new models.

Direct observational measurements of the gas content in distant
  galaxies are currently available for only a very limited number of
  galaxies, and are likely biased towards the most gas-rich objects. Gas
  is usually detected using CO molecules as a tracer, so no direct
  information about the atomic gas content of high redshift galaxies
  is currently available. In addition, the conversion from CO to
  molecular hydrogen mass is notoriously uncertain and may depend on
  galaxy properties \citep[][and references therein]{Genzel2010}. In the coming decade, it is hoped that new
  facilities like {\sc ALMA} (Atacama Large Millimeter array) and the
  {\sc SKA} (Square Kilometer Array) will reveal the gas content in
  representative samples of high-redshift galaxies. 

In the meantime, we can obtain {\em indirect} constraints on gas
content by using the observational estimates of galaxy size and SFR
and empirical correlations between SFR density and gas density. This
approach has been used by \citet{Erb2006} and \citet{Mannucci2009},
who obtained estimates of gas masses for high redshift galaxies by
inverting the Schmidt-Kennicutt relation (KS law) described in
Kennicutt (1998b), which relates SFR surface density to the combined
atomic and molecular hydrogen surface density. However, the indirect
estimates of gas fraction that they obtain do not agree well with the
direct measurements, as we later show. This could be because the
direct estimates are biased high, or it could be that the ``total
gas'' form of the KS law does not apply. It has been shown that the KS
law breaks down at the lowest gas surface densities
\citep{Bigiel2008}. This breakdown is ascribed to the inability of the
cold gas at low surface densities to collapse gravitationally and form
molecular clouds from which stars may originate. SFR surface density
correlates in an almost linear fashion with the molecular gas surface
density without breaking down at lower surface densities
\citep{Bigiel2008,Bigiel2011,Schruba2011}. Using an inverted
{\em molecular-gas-based SF law} therefore seems more appropriate to probe
the cold gas content of high-redshift galaxies. Furthermore, inverting
a molecular-gas-based SF law allows for comparison between directly
and indirectly measured molecular gas masses of high-redshift
galaxies. 

In this paper we present an improved method to derive the gas
properties of distant galaxies from their star formation rate
densities.
An integral part of our new method is a prescription to calculate
the molecular fraction of cold gas based on the results by
\citet{Blitz2006}. These authors found that the ratio between the
molecular and atomic hydrogen surface density ($R_{\mathrm{H}_2} =
\Sigma_{\mathrm{H}_2}/\Sigma_{\mathrm{HI}}$) in discs can be
described empirically as a function of the hydrostatic mid-plane
pressure $P_m$, driven by the stellar and gas density \citep[see
also][]{wong2002, blitz2004}. This allows us to invert the much
tighter and more physical molecular gas based star formation
relation derived for nearby galaxies by \citet{Bigiel2008}, instead
of the traditional KS relation.

We apply our method to $\sim57000$ galaxies from the \cosmos survey in
the redshift range of $0.5\leq z \leq 2.0$ to study the structure and
properties of gas and the galaxy evolution over this important cosmic
epoch. \cosmos was chosen for our study because it has well
established disc sizes based on high resolution (Hubble Space
Telescope) imaging, crucial for our method, imaging running from the
X-ray to radio wavelengths to derive stellar masses and SFR, and
spans a wide range of redshift.

The outline of this paper is as follows. In Section \ref{sec:method},
we present our method to indirectly measure the molecular and the
total cold gas content in galaxies.  In Section \ref{sec:calibration},
we apply this method to both low- and high-redshift data with known
gas properties and calibrate our results against direct measures of
cold and molecular gas. In Section \ref{sec:cosmos}, we describe the
application of the developed method to $\sim57,000$ galaxies from the
\cosmos survey in the redshift range of $0.5\leq z \leq 2.0$, and the
resulting gas properties with respect to other galaxy parameters. We
discuss our findings in Section \ref{sec:discussion} and summarize our
results in Section \ref{sec:summary}. Throughout the paper we apply a
$\Lambda$CDM cosmology with $H_0\,=\,70$ km s$^{-1}$,
$\Omega_{\mathrm{matter}}\,=\,0.28$, and $\Omega_\Lambda\,=\,0.72$. We
assume a universal Chabrier stellar initial mass function
\citep[IMF:][]{chabrier2003} and where necessary convert all
observations used to a Chabrier IMF. All presented cold and molecular
gas surface densities and integrated masses include a correction of
1.36 to account for Helium.

\begin{table*}
\caption{Summary of the parameters used in our indirect gas measure
  method\label{tab:parameters}. These parameters lead to the overall
  best fit results for integrated masses in Section \ref{sec:calibration}}
\begin{tabular}{llcl}
\hline
Parameter & Description & Value & Reference\\
\hline
$A_{\mathrm{SF}}$ & Normalization of the star formation law
($M_\odot\,\mathrm{yr}^{-1}\,\mathrm{kpc}^{-2}$) & $1.09\times10^{-2}$
& Sec. \ref{sec:changing_parameters} \\
$N_{\mathrm{SF}}$ &Index which sets star formation efficiency & 0.5 &
Sec. \ref{sec:changing_parameters}\\
$\Sigma_{\mathrm{crit}}$ & Critical surface density which sets
star formation efficiency ($M_\odot\,\mathrm{pc}^{-2}$) & 100 & Sec. \ref{sec:changing_parameters}\\
$P_0$ & Normalization of the \citet{Blitz2006} power-law & $2.35\times
10^{-13}$ & \citet{Leroy2008}\\
$\alpha$ & Index of the \citet{Blitz2006} power-law & 0.8 &
\citet{Leroy2008}\\
$ \bar f_\sigma$ & Mean ratio between gas and stellar vertical
velocity dispersion in the stellar disc & 0.4 &
\citet{Elmegreen1993}\\
$\chi_{\mathrm{gas}}$ & Scale radius of the gas disc, relative to the
stellar disc & 1.5 & Sec. \ref{sec:changing_parameters}\\
\hline
\hline
\end{tabular}
\end{table*}

\section{Method}
\label{sec:method}
\subsection{Calculating gas surface densities}
We infer gas surface densities using a combination of an empirical
molecular star formation law \citep[based on][]{Bigiel2008} and a
prescription to calculate the fraction of molecular hydrogen in cold
gas \citep[based on][]{Blitz2006}.

We have slightly adapted the star formation law deduced by
\citet{Bigiel2008} to allow for higher star formation efficiencies in
high gas surface density regions. This is based on the results of
\citet{Daddi2010} and \citet{Genzel2010}, who found the star-formation
at high surface densities to follow the KS - law (power-law slope of
1.4 versus 1.0 for \citet{Bigiel2008}). The resulting equation is given by
\begin{equation}
\Sigma_{\mathrm{SFR}} = \frac{A_{\mathrm{SF}}}{10 M_\odot\,\mathrm{pc}^{-2} }\,\left(1 + \frac{\Sigma_{\mathrm{gas}}}{\Sigma_{\mathrm{crit}}}\right)^{N_{\mathrm{SF}}}\,f_{\mathrm{H}_2}\,\Sigma_{\mathrm{gas}}
\end{equation}
where $\Sigma_{\mathrm{SFR}}$ and $\Sigma_{\mathrm{gas}}$ are the star
formation and cold gas surface densities in
$M_\odot\,\mathrm{yr}^{-1}\,\mathrm{kpc}^{-2}$ and $M_\odot\,\mathrm{pc}^{-2}$,
respectively, $A_{\mathrm{SF}}$ is the normalization of the power law in
$M_\odot\,\mathrm{yr}^{-1}\,\mathrm{kpc}^{-2}$, $\Sigma_{\mathrm{crit}}$ a
critical surface density above which the star formation follows
\citet{Kennicutt1998law}, $N_{\mathrm{SF}}$ is an index which sets the
efficiency, and
$f_{\mathrm{H}_2}=\Sigma_{\mathrm{H}_2}/(\Sigma_{\mathrm{HI}}+\Sigma_{\mathrm{H}_2}$
is the molecular gas fraction. Table \ref{tab:parameters} gives a
summary of the parameter values and their origin.

Given a star-formation surface density, this equation can be solved
for the cold gas surface density when one knows the molecular fraction
of the cold gas. We use a pressure-regulated recipe to determine the
molecular fraction of the cold gas, based on the work by
\citet{Blitz2006}. They found a power-law relation between the
mid-plane pressure acting on a galaxy disc and the ratio between
molecular and atomic hydrogen, i.e.,
\begin{equation}
R_{\mathrm{H}_2} = \bigl(\frac{\Sigma_{\mathrm{H}_2}}{\Sigma_{\mathrm{HI}}}\bigr) = \bigl(\frac{P_m}{k_BP_0}\bigr)^\alpha
\label{eq:blitz2006}
\end{equation}
with $P_0$ the external pressure in the interstellar medium where the
molecular fraction is unity. $\alpha$ and $k_B$ are the power-law
index and the Boltzmann constant, respectively (see Table
\ref{tab:parameters}). $P_m$ is the mid-plane pressure acting on the
galaxy disc, and is given by \citep{Elmegreen1989}
\begin{equation}
P_m(r) = \frac{\pi}{2}\,G\,\Sigma_{\mathrm{gas}}(r)\left(\Sigma_{\mathrm{gas}}(r) + f_{\sigma}(r)\Sigma_*(r)\right)
\label{eq:pressure}
\end{equation}
where G is the gravitational constant, $r$ is the radius from the
galaxy centre, and $f_\sigma(r)$ is the ratio between
$\sigma_{\mathrm{gas}}(r)$ and $\sigma_*(r)$, the gas and stellar vertical
velocity dispersion, respectively.  We followed the method presented
in \citet{Fu2010} to determine $f_\sigma$ as a function of radius
$r$. Observations have shown that $\sigma_{\mathrm{gas}}$ remains constant
over the disc \citep[e.g.][]{shostak1984,Leroy2008}. The stellar
vertical velocity dispersion $\sigma_*$ decreases exponentially and
has a scale length twice the stellar disc scale length $r_*$
\citep[e.g.][]{Bottema1993}, i.e.
\begin{equation}
\sigma_*(r) = \sigma_*^0\exp{(-r/2r_*)}
\end{equation}
where the stellar velocity dispersion in the center of the disc is
noted as $\sigma^0_*$. The stellar mass surface density of an
exponential disc is given by $\Sigma_* = \Sigma^0_*\exp{(-r/r*)}$,
which we can combine with the previous equation to find
\begin{equation}
f_\sigma(r) =
\frac{\sigma_{\mathrm{gas}}}{\sigma_*^0}\sqrt{\frac{\Sigma_*^0}{\Sigma_*(r)}}
\equiv f^0_\sigma \sqrt{\frac{\Sigma_*^0}{\Sigma_*(r)}}
\label{eq:fratio}
\end{equation}
where $f^0_\sigma$ is the ratio between the gas and stellar vertical
velocity dispersion at the centre of the disc and $\Sigma_*^0$ is the
stellar mass surface density in the centre of the disc. We can express
$f_\sigma^0$ in terms of the mean value of $f_\sigma(r)$ in the disc
by
\begin{equation}
\bar f_\sigma = \frac{\int 2 \pi r \Sigma_*(r) f_\sigma(r) dr}{\int 2\pi r \Sigma_*(r) dr}
\end{equation}
Substituting for $f_\sigma(r)$ and afterwards integrating
we find
\begin{equation}
\bar f_\sigma  = 4 f_\sigma^0.
\end{equation}
By combining Equation (\ref{eq:pressure}) and Equation 
(\ref{eq:fratio}) we find
\begin{equation}
P_m(r) =
\frac{\pi}{2}\,G\,\Sigma_{\mathrm{gas}}(r)\bigl(\Sigma_{\mathrm{gas}}(r) + \frac{\bar f_\sigma}{4}\sqrt{\Sigma_*(r)\Sigma_*^0}\bigr)
\end{equation}
We now have all the necessary ingredients to calculate
$R_{\mathrm{H}_2}$ and subsequently the cold gas molecular fraction
[$f_{\mathrm{H}_2}=R_{\mathrm{H}_2} /(1+R_{\mathrm{H}_2})]$.

Putting this together, we have an expression for the star formation
law depending on the cold gas surface density, stellar mass surface
density and stellar mass and scale length of the disc. Given a star
formation surface density and stellar surface-density, we can now
indirectly determine the cold gas surface density through
iteration. The molecular gas fraction, and subsequently the molecular
gas surface density, can then easily be calculated from Equation
(\ref{eq:blitz2006}), using the stellar mass surface density and
previously determined gas surface density as input.

Intrinsically hidden in this method is the CO-to-\h2 conversion factor
($X_{\mathrm{CO}}$). This conversion factor was used to determine 1)
the normalization constant $A_{\mathrm{SF}}$ for the empirical star
formation law based on \h2 \citep{Bigiel2008} and 2) the ratio between
molecular and atomic hydrogen $R_{\mathrm{H}_2}$ which was used to
determine the normalization and power-law index for the pressure-based
\h2 fraction recipe \citep{Leroy2008}. However, \citet{Blitz2006}
proposed that changes in the CO-to-\h2 conversion factor are unlikely
to substantially alter results for the latter point. Unless stated
differently, we adopted a value of $X_{\mathrm{CO}} = 2.0\times10^{20}
\,\mathrm{cm}^{-2}\,(\mathrm{K\,km\,s}^{-1})^{-1}$. 

\begin{figure}
\includegraphics[width = 1.0\hsize]{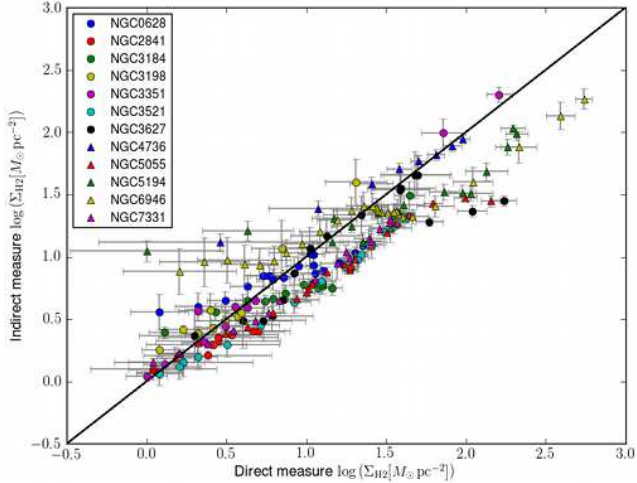}
\caption{Comparison between indirect and direct measures of
  molecular gas surface densities, taken from the radial profiles of
  the THINGS galaxies \citep{Leroy2008} using our
  method. \label{fig:1_1_THINGS_H2} All datapoints are coded by galaxy
  and represent surface densities sampled along their host galaxy
  radius.}
\end{figure}

\subsection{Constructing a galaxy}

The method described in the previous subsection allows one to
calculate the gas surface density, given a measured stellar-mass
surface density and SFR surface density. However, for many
high-redshift objects only global parameters are available, combined
with some morphological information (e.g., stellar scale length $r_*$,
half light radius $r_{0.5}$).

In order to calculate the cold and molecular gas content of these
galaxies, we distribute the stellar mass into an exponential disc
following $\Sigma_*(r) = \Sigma_*^0\exp(-r/r_*)$ where $r_*$ is the
scale length of the stellar disc and $\Sigma_*^0$ is the stellar mass
surface density in the centre of the disc (defined as $\Sigma_*^0 =
M_*/(2\pi r_*^2)$ where $M_*$ is the galaxy stellar mass). The star
formation in the galaxy is distributed similarly, with a disc scale
length
\begin{equation}
r_{\mathrm{gas}} = \chi_{\mathrm{gas}} \,r_*,
\end{equation}
where $\chi_{\mathrm{gas}}$ is a scale radius of the gas disc relative
to the stellar disc (see Table \ref{tab:parameters}). The use of
$\chi_{\mathrm{gas}}$ is motivated by the radial extent of \hi gas in
spiral galaxies \citep{broeils1997}. A galaxy's total
cold gas budget can be found by integrating over the inferred gas
surface densities. We find the best fit results for integrated masses
(see Section \ref{sec:galaxy_comparison}) when distributing the
stellar mass and star formation over an exponential disc out to five
times the stellar disc scale length.

\begin{figure*}
\includegraphics[width = 1.0\hsize]{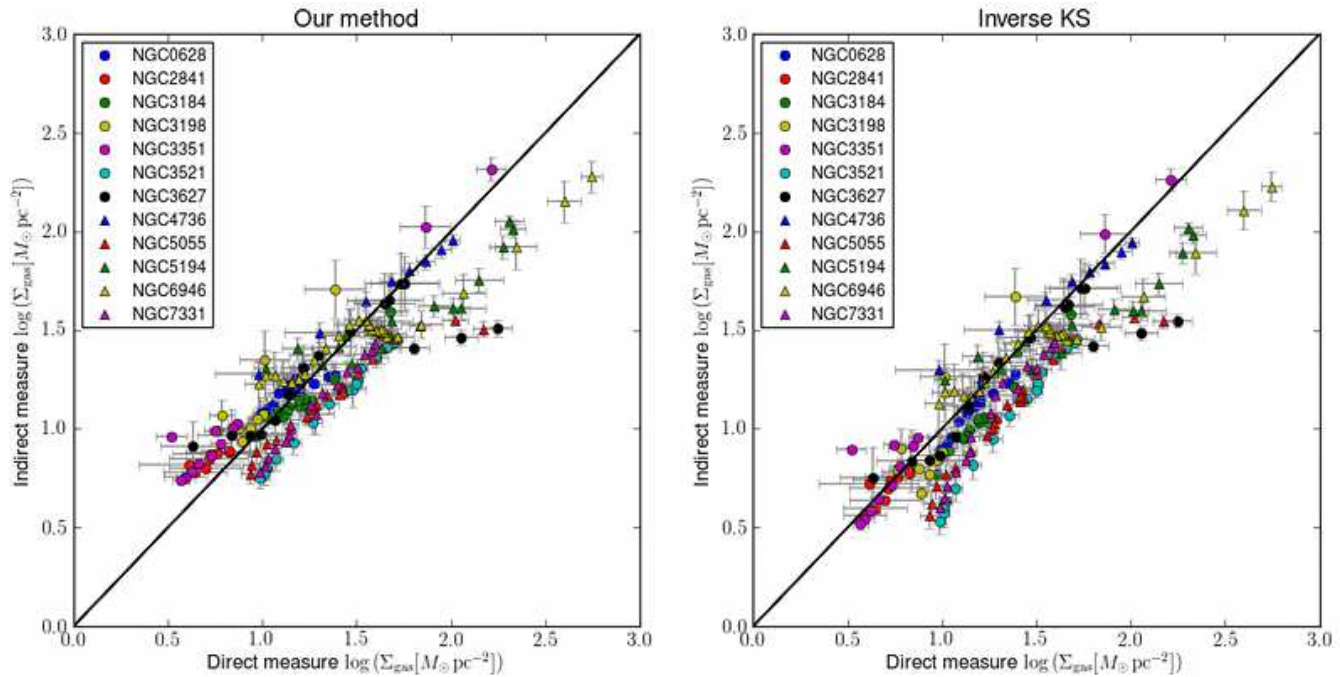}
\caption{Indirect versus direct measures of total cold gas surface
  densities, taken from the radial profiles of the THINGS galaxies
  \citep{Leroy2008}. Left and right panel using our and the KS
  indirect gas measure method, respectively. We find better agreement
  with literature values than the indirect KS method, especially at
  low surface densities.  \label{fig:1_1_THINGS_gas}}
\end{figure*}

\section{Calibration of the method}
\label{sec:calibration}
In this section we compare indirect measures of gas surface densities
and integrated masses obtained using the method presented in the
previous section to direct measures from the literature. Results were
obtained using the parameter values as summarized in Table
\ref{tab:parameters}. Note that the purpose of this method was to
measure the global cold gas and molecular content of galaxies, rather than
local surface densities. Parameters were chosen to best achieve this
purpose. We discuss the effects of changing the parameters in Section
\ref{sec:changing_parameters}.

\subsection{Local galaxy surface densities}
Figures~\ref{fig:1_1_THINGS_H2} and \ref{fig:1_1_THINGS_gas} show
total cold gas and molecular gas surface densities obtained using our
method. The indirect measures were computed from the radial profiles
of the star-formation surface densities and stellar mass surface
densities, in combination with the host galaxy total mass and stellar
scale length all obtained from \citet{Leroy2008}.  These predictions
are plotted against literature values for total cold gas and molecular
gas surface densities taken from \citet{Leroy2008}

We find good general agreement between results obtained using our
method and literature values at molecular surface densities ranging
from $\Sigma_{\mathrm{H}_2} \sim 1$ to $\Sigma_{\mathrm{H}_2} \sim
300\, M_\odot\,\mathrm{pc}^{-2}$
(Figure~\ref{fig:1_1_THINGS_H2}). There is a scatter around the
one-to-one line of $0.1$ to $0.3$ dex at the lowest and more
intermediate surface densities, respectively. We overestimate the
lowest molecular surface densities for NGC 6946 and NGC 5194, whereas
we under-predict at high surface densities for these galaxies (also
see NGC 3627). The star-formation efficiency (SFE, the ratio of SFR over
molecular gas) of these galaxies decreases significantly towards
smaller radii (at the highest surface densities) compared to the
predicted SFE from the mid-plane pressure on the disc \citep[see
  Appendix F of][]{Leroy2008}. The low SFE of these galaxies with
respect to theoretical values will result in the observed surplus of
molecular gas as predicted by our method. Opposite behavior is
observed for these galaxies towards larger radii, where the molecular
surface density is much lower. In this case, deviations between
observed and theoretical SFE result in a deficit of molecular hydrogen as
computed by our method.

We also find very good agreement between the total gas surface
densities computed using our method and surface densities
obtained from the literature (left-hand panel of Figure
\ref{fig:1_1_THINGS_gas}), with smaller scatter especially at the
lower gas surface densities. We predict a deficit of total cold gas
compared to literature values at the highest surface
densities. Similar to the molecular gas surface densities, we ascribe
this deficit to differences between theoretically predicted and
observed SFE. Overall we find a mean ratio of 
$\Sigma_{\mathrm{gas}\,\mathrm{indirect}}/\Sigma_{\mathrm{gas}\,\mathrm{ observed}} = 0.94$
with a standard deviation of 0.021 for our method.

We plot indirect total cold gas surface densities obtained with the KS
law as a function of the literature values in the right-hand panel of
Figure~\ref{fig:1_1_THINGS_gas} for comparison. Although the scatter
around the one-to-one line of the two methods is similar, there is a
clear offset from this line when using the KS law, resulting in a
general deficit of gas over the entire range of total gas surface
densities. We find a mean ratio $\Sigma_{\mathrm{gas}\,\mathrm{indirect}}/\Sigma_{\mathrm{gas}\,\mathrm{observed}} = 0.79$
with a standard deviation of 0.019 when using the KS law for the
indirect gas measure. This was expected for the lowest surface densities where
the KS law breaks down \citep{Bigiel2008}, which also appears in
Figure \ref{fig:1_1_THINGS_gas}. However, the offset is surprising at
the higher gas surface densities at which the KS law is defined to
relate SFR surface densities with the total gas surface
density. Putting the observed gas surface densities in $\Sigma_{\mathrm{SFR}}-\Sigma_{\mathrm{gas}}$ space and
comparing with the KS law shows that the data lie slightly below the
KS law. Nevertheless, this can not account for the breakdown at low gas
surface densities. We therefore find that our method yields the biggest improvements
at the lowest gas surface densities (i.e., the outer parts of galaxy
discs). 

\begin{figure}
\includegraphics[width = \hsize]{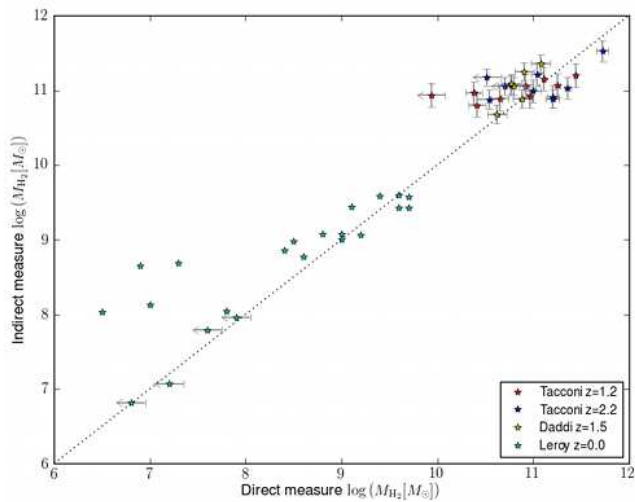}
\caption{Indirect vs. direct measures of galaxy molecular gas masses
  using our indirect gas method. Both low- \citep{Leroy2008} and high-
  \citep{Tacconi2010, Daddi2010} redshift galaxies are
  displayed. Apart from four local dwarf galaxies (see
  Sec.~\ref{sec:galaxy_comparison}), we find good agreement between
  literature values and our method at all redshifts and molecular gas
  masses. \label{fig:one_to_one_galaxy_H2}}
\end{figure}

\subsection{Galaxy integrated gas masses}
\label{sec:galaxy_comparison}

\begin{figure*}
\includegraphics[width = \hsize]{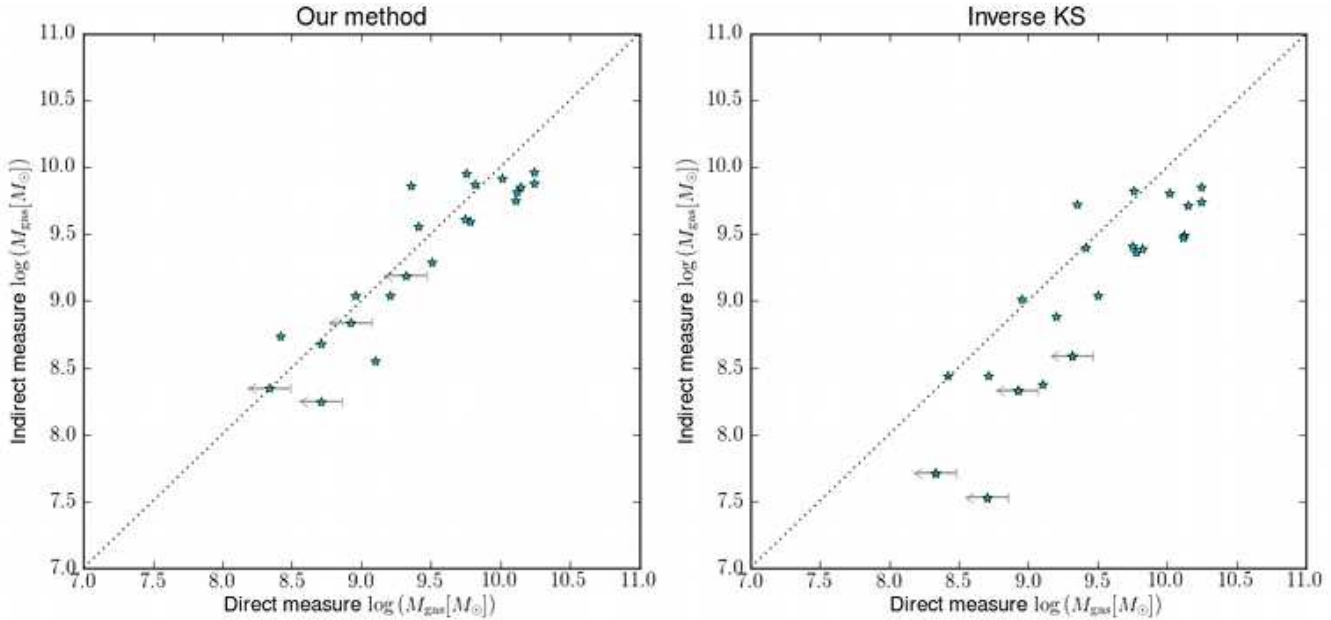}
\caption{Indirect vs. direct measures of galaxy cold gas masses using 
  our (left panel) and the KS (right panel) indirect gas
  method. No neutral gas measurements are available for the
  high-redshift galaxies, therefore only low-redshift galaxies are
  plotted. Our method finds good agreement with literature values,
  whereas the inverse KS method find a deficit of total cold gas
  mass.\label{fig:one_to_one_galaxy_gas}}
\end{figure*}

We present galaxy integrated \h2 and total cold gas masses obtained
with our method versus literature values in Figures
\ref{fig:one_to_one_galaxy_H2} and \ref{fig:one_to_one_galaxy_gas},
respectively. We applied our method to a small sample of both low and
high-redshift galaxies, compiled from \citet[$z=0$]{Leroy2008},
\citet[$z\sim 1.5$]{Daddi2010}, and \citet[$z\sim1.2$ and $z\sim
  2.2$]{Tacconi2010}. We applied the $X_{\mathrm{CO}}$ factors as
determined by the latter two authors for their galaxy samples
[$X_{\mathrm{CO}} \sim 2.5\times10^{20}$ and $X_{\mathrm{CO}} =
  2.0\times10^{20}\,\mathrm{cm}^{-2}\,(\mathrm{K\,km\,s}^{-1})^{-1}$,
  respectively].

We find good agreement between our results and observations over the
entire range of \h2 masses and redshifts. There is a slight excess of
\h2 computed with our method as compared to the direct measurements
($\sim 0.2$ dex; Figure \ref{fig:one_to_one_galaxy_H2}). We find a
spread of $\sim 0.25$ dex around the mean trend in computed versus
literature \h2 masses. We neglect the four local galaxies with a
surplus of integrated \h2 masses of $1$--$1.5$ dex. All four are
classified as dwarf galaxies and have anomalously high SFE in their
central parts \citep{Leroy2008}. These authors argue for an
unaccounted-for reservoir of \h2 in these galaxies, possibly due to
variations in the $X_{CO}$ factor for dwarfs with respect to bigger
spiral galaxies \citep[see also][]{spaans1997}. We find one clear
outlier at higher redshift, with an upper limit on its observed \h2
mass. It is most likely that this galaxy also has a high SFE we cannot
account for.

Current instruments do not allow for atomic hydrogen observations at
high redshifts. We are therefore restricted to local galaxies when
comparing literature values to total integrated galaxy gas masses
computed with the method presented in this study. We find good
agreement between our results and the observational data, although
with a slight mean deficit (left-hand panel of Figure
\ref{fig:one_to_one_galaxy_gas}). We found a maximum deviation in
literature values and our results of $\sim 0.5$ dex. Predicted cold
gas masses for most galaxies are within $\sim 0.3$ dex of their
literature values. It is worth noting that the four dwarfs for which
we over-predict integrated \h2 masses are also included in this figure
and have predicted total gas masses in good agreement with the
observations. Although this seems promising, it means that not only do
we find a surplus of \h2 compared to current literature studies, we
also find a deficit of atomic hydrogen.

In the right hand panel of Figure \ref{fig:one_to_one_galaxy_gas} we
present indirect total cold gas measures for local galaxies obtained
by applying the KS law. Similar to the surface density profiles, the
KS law predicts a deficit of total integrated gas mass. The outer
parts of the galaxies (where the gas surface densities are lowest, the
KS law breaks down and more gas is necessary to sustain SF) drive
this deficit. Here we clearly see the strength of our method over
previous indirect gas methods.

These results demonstrate out that our method can be used as a useful tool
to predict the cold and molecular gas surface densities, as well as
integrated gas masses for low- and high-redshift galaxies.

\subsection{Changing parameters}
\label{sec:changing_parameters}
The method presented in this paper has several free parameters, which
all individually affect the outcome of indirect gas measures. In this
subsection we explore the spread in results due to variations in free
parameters.

We find that our method is robust against changes in the mean ratio
between gas and stellar vertical velocity dispersion $\bar
f_\sigma$. We explore values ranging from $0.1$ to $2.0$ and find
differences in total cold gas surface density measures up to $\sim0.1$
dex with respect to results obtained using the quoted value for $\bar
f_\sigma$ in Table (\ref{tab:parameters}). Differences are most
prominent at the lowest gas surface densities ($\Sigma_{\mathrm{gas}}
\leq 10 M_\odot\,\mathrm{pc}^{-2}$). The spread in integrated total
cold gas masses due to changes in $\bar f_\sigma$ also reaches only
$\sim$0.1 dex over the entire mass range probed by our control sample
of galaxies. Changing $\bar f_\sigma$ does not significantly affect
molecular gas surface densities and integrated masses.

\citet{Blitz2006} derived a power-law index $\alpha$ and a
normalization constant $P_0$ slightly different than those we use
($\alpha$ = 0.8 versus 0.92 and $P_0 = 5.93\times 10^{-13}$ versus
$2.35\times 10^{-13}$, respectively). Applying their values to our
method results in small changes of only $\sim0.1$ dex at lower gas
surface densities with respect to indirect gas measures obtained using
the parameters as summarized in Table~\ref{tab:parameters}. Integrated
gas masses differ within $\sim0.15$ dex over the entire mass range
probed.

The most significant changes in inferred indirect gas measures are due
to variations in the normalization of the star formation law
$A_{\mathrm{SF}}$. In recent works \citet{Bigiel2008} and
\citet{Bigiel2011} found values for the star formation normalization
of $A_{\mathrm{SF}} = 8.42\times10^{-3}
M_\odot\,\mathrm{yr}^{-1}\,\mathrm{kpc}^{-2} $ and $A_{\mathrm{SF}} =
4.6\times 10^{-3} M_\odot\,\mathrm{yr}^{-1}\,\mathrm{kpc}^{-2}$,
respectively (after conversion to a Chabrier IMF). We explored a range
of parameters around these values and applied a $\chi^2$ test on the
derived integrated molecular and atomic gas masses (excluding dwarf
galaxies). This resulted in a best fit value of $A_{\mathrm{SF}} =
8.4\times 10^{-3} M_\odot\,\mathrm{yr}^{-1}\,\mathrm{kpc}^{-2}$ for
the integrated total cold gas masses, and $A_{\mathrm{SF}} =
13.5\times 10^{-3} M_\odot\,\mathrm{yr}^{-1}\,\mathrm{kpc}^{-2}$ for
the integrated molecular gas masses. We found that $A_{\mathrm{SF}} =
10.9\times 10^{-3} M_\odot\,\mathrm{yr}^{-1}\,\mathrm{kpc}^{-2}$ leads
to best overall agreement between direct and indirect measures for
both the total cold and molecular gas integrated masses (see
Table~\ref{tab:parameters}).

Adopting other values (running from $A_{\mathrm{SF}} = 1.0\times
10^{-3}$ to $A_{\mathrm{SF}} = 9.0\times
10^{-4} M_\odot\,\mathrm{yr}^{-1}\,\mathrm{kpc}^{-2}$) for the normalization of the star formation law
leads to differences of a few tenths of a dex between literature
values and our total cold gas and molecular gas results. For example,
$A_{\mathrm{SF}} = 4.6\times 10^{-3}
M_\odot\,\mathrm{yr}^{-1}\,\mathrm{kpc}^{-2}$ \citep{Bigiel2011} results in total gas
surface densities differing by $\sim 0.2$ dex from results obtained
using the parameter values presented in
Table~\ref{tab:parameters}. This is a systematic shift upwards in
indirect surface density over the entire range of surface densities
probed. Changes in integrated indirect mass measures are of the order
$\sim0.15$ dex with respect to results obtained using the tabulated
parameter values.

Offsets between molecular gas direct and indirect measures are more
prominent, up to 0.5 dex for the integrated molecular gas masses when we
adopt $A_{\mathrm{SF}} = 4.6\times 10^{-3}
M_\odot\,\mathrm{yr}^{-1}\,\mathrm{kpc}^{-2}$. Adopting
$A_{\mathrm{SF}}$ with values larger than presented in Table
\ref{tab:parameters} results in a similar systematic shift downwards
for predicted surface densities and integrated masses.

We applied an adapted formalism of the star formation law presented in
\citet{Bigiel2008} to allow for higher SFE in dense regions
\citep[motivated by][]{Daddi2010,Genzel2010}. Adopting instead the
standard SFE has no significant influence on the integrated cold gas and
molecular gas predictions of local galaxies. Less concordance is
reached between literature values and our predictions for the high
redshift objects. We find a shift of approximately $\sim 0.2$ dex
upwards away from the one-to-one line with respect to results obtained
with the inclusion of an increased SFE at highest gas surface densities.

One of the key assumptions in this method is an exponential
distribution of matter in the galaxy discs. SF occurs in molecular
clouds (local clumps in the disc not following an exponential
distribution), which could lead to a local underestimation of the
cold gas surface density. Nevertheless, an exponential disc seems a
valid approximation for star-forming galaxies on the main sequence
\citep{wuyts2011}, and the local clumps should average out when
integrating over the disc. As we will discuss in Section
\ref{sec:sample}, part of our \cosmos galaxy sample consists of
quenched galaxies which fall off the main sequence. We are aware that
these galaxies are not described by disks but that their light
profiles are better described by a Vaucouleurs profile \citep{wuyts2011}.

We assume that the gas in galactic discs is radially more extended than
the stars. Varying the scale length of gas relative to stars,
$\chi_{\mathrm{gas}} \geq 1$, results in only subtle changes in the
integrated total cold and molecular gas masses. However, decreasing
the gas scale length to $\chi_{\mathrm{gas}} < 1$ lowers the
inferred integrated cold (molecular) gas masses.  This difference
can increase to $\sim 0.25\, (0.1)$ dex when $\chi_{\mathrm{gas}} =
0.5$ with respect to results obtained using the parameters presented
in Table \ref{tab:parameters}.

Both the SFR and stellar mass are a function of the IMF. Changes in
inferred total cold gas surface densities are less than 0.1 dex at the
lowest gas surface densities when adopting a Salpeter IMF
\citep{salpeter1955} rather than the Chabrier IMF used above, and
these changes become negligible at the highest surface densities. The
integrated cold gas mass of the galaxies, as well as molecular gas
surface densities and integrated masses do not change significantly
when adopting a Salpeter IMF.

\begin{figure*}
\includegraphics[width = 1.0\hsize]{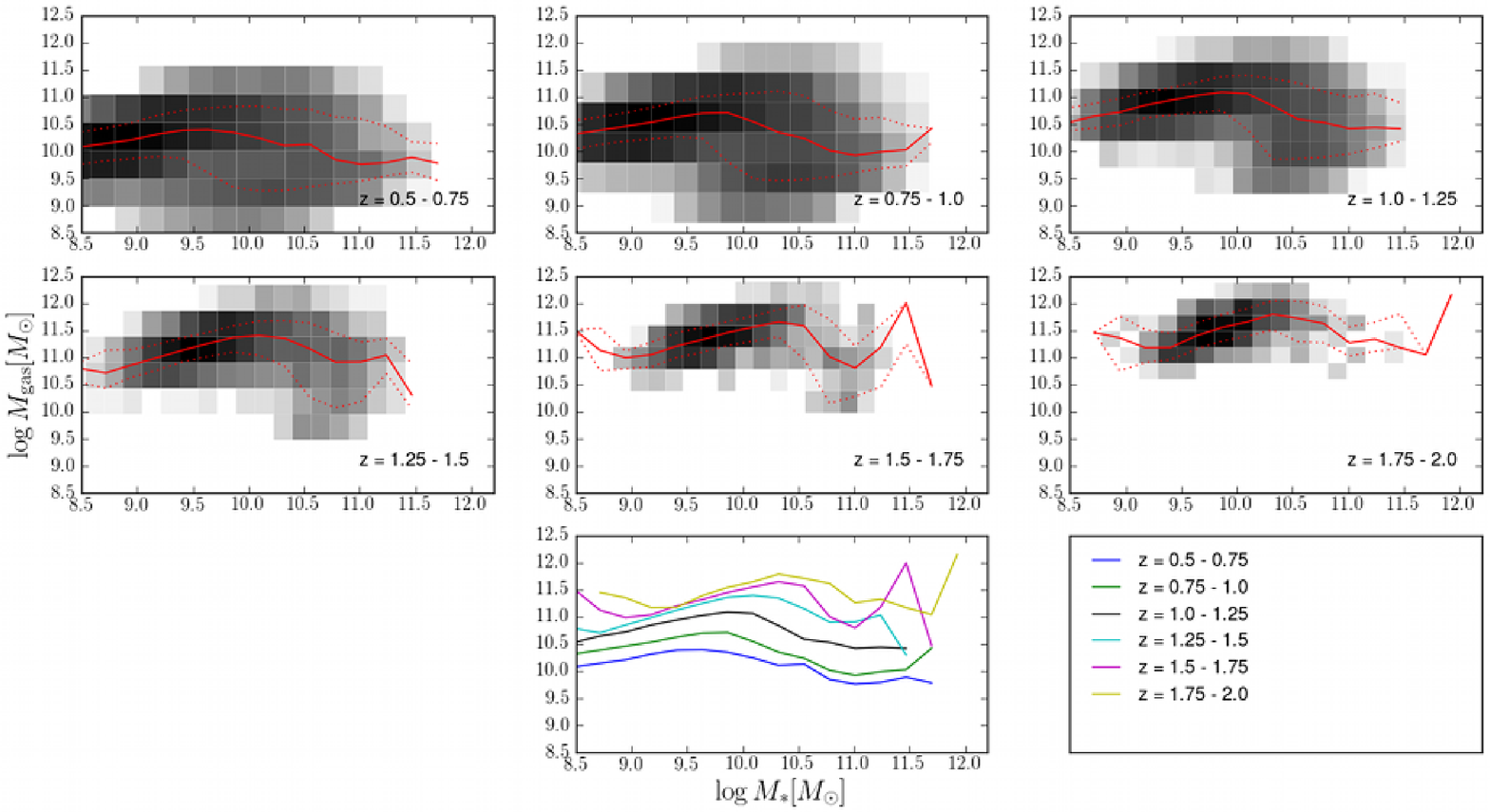}
\caption{Total-cold mass gas as a function of stellar mass for
  different redshift bins. The grey shaded area shows the log of the
  number of galaxies in each gas/stellar mass bin, with the 50, 16 and
  84 percentile curves shown with the red solid and dashed lines. The
  central bottom panel shows the 50 percentile curve for the data in
  each redshift bin. Bumps at the high-mass end in the highest
  redshift bins are due to ULIRGS (see Sec. \ref{sec:gas_content}\label{fig:cosmos_gas})}
\end{figure*}

\begin{figure*}
\includegraphics[width = 1.0\hsize]{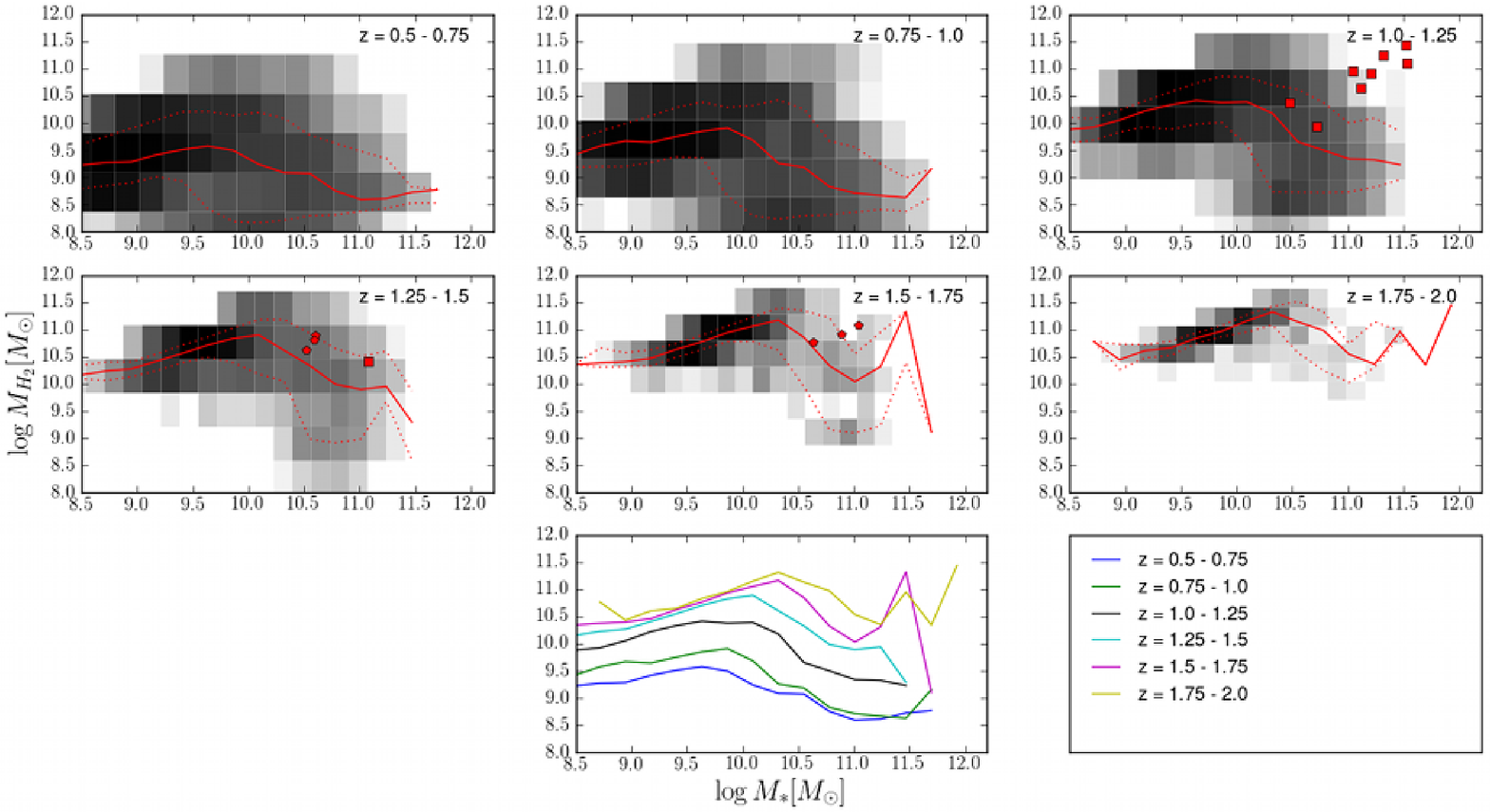}
\caption{\h2 mass as a function of stellar mass for different redshift
  bins. The grey shaded area shows the log of the number of galaxies
  in each bin, with the 50, 16 and 84 percentile curves shown with the
  red solid and dashed lines. The central bottom panel shows the 50
  percentile curve for the data in each redshift bin. Red squares and
  pentagons are literature values taken from \citet{Tacconi2010} and
  \citet{Daddi2010}, respectively. \label{fig:cosmos_H2}}
\end{figure*}

\begin{figure*}
\includegraphics[width = 1.0\hsize]{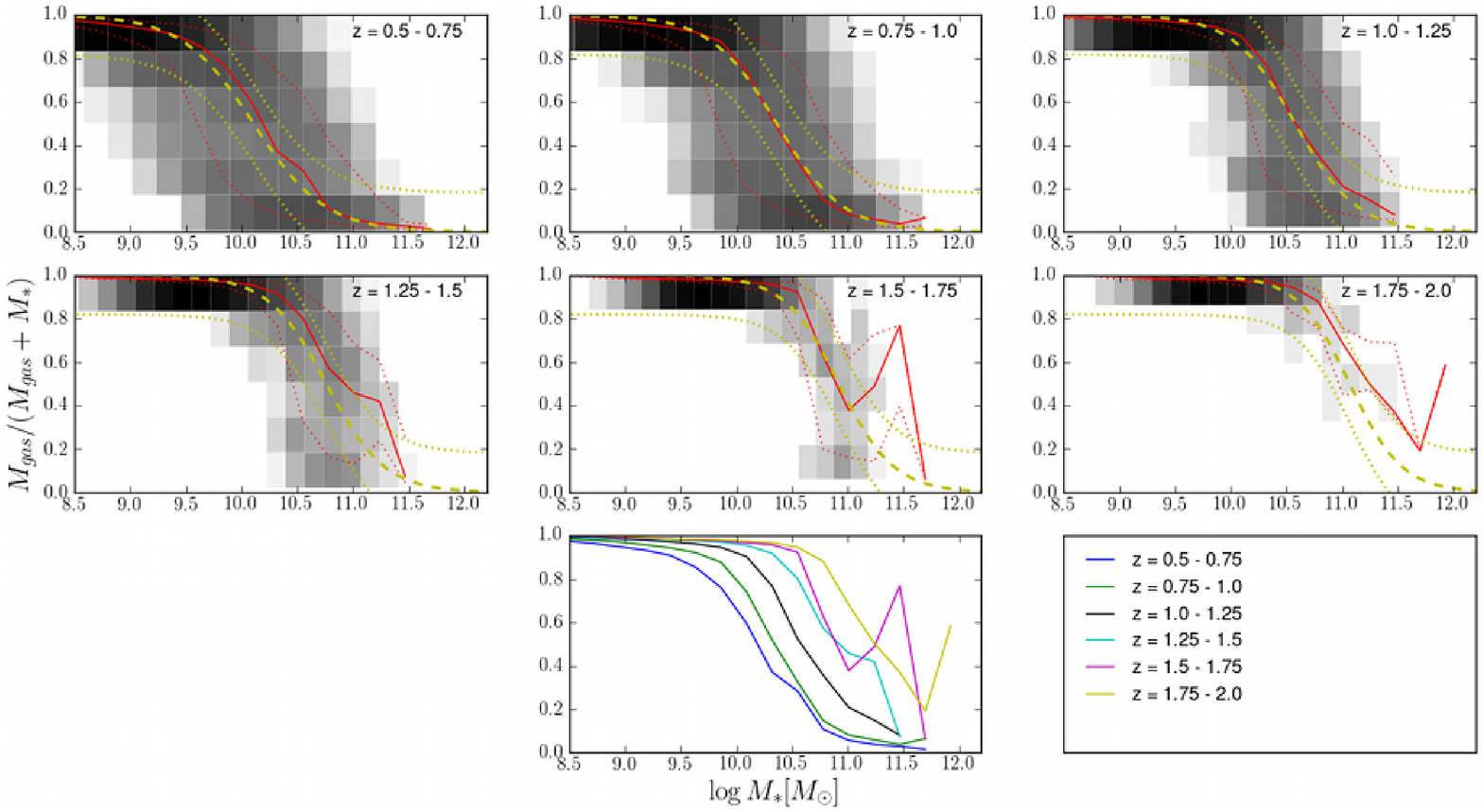}
\caption{Total cold gas fraction (atomic plus molecular) as a function
  of stellar mass for different redshift bins. The grey shaded area
  shows the log of the number of galaxies in each bin, with the 50, 16
  and 84 percentile curves shown with the red solid and dashed
  lines. The central bottom panel shows the 50 percentile curve for
  the data in each redshift bin. The dashed and dotted yellow lines
  represents the fit and one sigma scatter of Equation
  \ref{eq:gas_frac_fit}, respectively.\label{fig:gas_frac}}
\end{figure*}

\section{Gas evolution in the COSMOS sample}
\label{sec:cosmos}

\subsection{Sample}
\label{sec:sample}
In order to determine the cold gas content and atomic and molecular
composition of high-redshift galaxies, we used a galaxy sample taken
from the Cosmic Evolution Survey \citep[\cosmos;][]{scoville2007}. The
survey was designed to probe the evolution of galaxies, star
formation, active galactic nuclei, and dark matter over the redshift
range $z > 0.5 - 6$. These data comprise imaging and spectroscopy from
X-ray to radio wavelengths including HST imaging and cover a
two-square degree area on the sky.

We considered a catalogue of $\sim$ 57000 galaxies with HST {\it I}
-band magnitude {\it I}$_{AB}<$ 24 mag in the redshift range $0.5 <
z_{\mathrm{phot}}<2.0$ from the \cosmos survey. Photometric redshifts
were obtained from \citet{Ilbert2009}. SFRs are based on rest-frame UV
fluxes, corrected using reddening values E(B-V) obtained from
\citet{Ilbert2009} for each individual galaxy. UV fluxes were converted into SFR
following \citet{Kennicutt1998SFR} for a Chabrier-IMF. We computed
stellar masses using a multi-wavelength SED $\chi^2$ fitting to the
\cosmos photometry ({\it U} to {\it K}-band), applying the
\citet{bruzual2007} templates fixed at the photometric redshifts of
the sample objects. We used galaxy half light radii ($r_{50}$) taken
from \citet{Scarlata2007} and \citet{Tasca2009}.

The selection of galaxies in the {\it I}-band does not necessarily guarantee that all selected galaxies lie on the
M$_*$--SFR plane. Our selection rather consists of both active star
forming galaxies and galaxies already quenched. Inspection of the
data points out that quenched galaxies indeed are especially dominant towards the
massive end of our galaxy sample ($\log{M_*} > 11\, [M_\odot]$).

We warn the reader that the sampling of galaxies at $z>1.5$ (and
especially $z>1.75$) is sparse, particularly at the low-mass end and one should use caution interpreting
results in this redshift range.

\subsection{Cold gas content}
\label{sec:gas_content}
We present the derived total cold gas and molecular gas masses in
Figures \ref{fig:cosmos_gas} and \ref{fig:cosmos_H2}. Grey scales
represent the log of the number of galaxies in each bin in the
$M_{\mathrm{gas}}\,(M_{H2})$--$M_*$ plane, with 50, 16 and 84
percentiles over-plotted on the figure. Results are presented as a
function of stellar mass in different redshift bins. Cold gas masses
range from $\log{M_{\mathrm{gas}}} \sim 10$--$10.5\,M_\odot$ at $z =
0.5$--$0.75$ with a peak around $\log{M_*}\sim \, 9.5\,M_\odot$ to
$\log{M_{\mathrm{gas}}} \sim 11$--$12\,M_\odot$ at the highest
redshifts in our sample peaking at a stellar mass of $\log{M_*}\sim
\,10.3\,M_\odot$.

We find a large spread in molecular gas masses in the range of
$\log{M_{H2}} \sim 8.0$--$11.0\,M_\odot$ at the lowest redshifts in
our sample and $\log{M_{H2}} \sim 10.0$--$11.5\,M_\odot$ for the
highest-redshift objects in our sample. The molecular gas masses peak
at the same stellar masses as the integrated total cold gas masses.

Not only do the cold and molecular gas masses of galaxies at fixed
stellar mass decrease with time, the largest gas reservoirs are always
found in less- (stellar) massive objects. This trend suggests that
(stellar) massive galaxies consumed their cold gas first, and only at
later stages do less-massive galaxies fully consume their gas
reservoir. We have explored the SFRs of galaxies of our sample
galaxies as a function of redshift and found that for a given stellar
mass, higher-redshift objects have higher SFR (i.e., consume their gas
faster) than their low-redshift counterparts.

We have superimposed observational results obtained by
\citet{Daddi2010} and \citet{Tacconi2010} on Figure
\ref{fig:cosmos_H2}. We obtain good agreement between molecular gas
masses computed with our method and observations extracted from the
literature. Only in the redshift regime $z = 1.0$--$1.25$ do we
compute a deficit of \h2 compared with results of
\citet{Tacconi2010}. 

As pointed out in section \ref{sec:sample}, quenched galaxies make
up for the dominant part of the massive objects in our sample and do
not track the M$_*$--SFR plane. The \citet{Tacconi2010} galaxies on the other hand are actively forming
stars and lie on the M$_*$--SFR plane. We find that these
literature galaxies have specific SFR (SSFR) much higher than
galaxies in the same redshift bin and with similar stellar masses
in our \cosmos sample ($>1.0$ dex). This difference in galaxy 
population (and consequently SSFR) can explain for the surplus of \h2 in literature
galaxies compared to our predictions.


We find a few massive objects
($\log{M_*}\sim11.5$--$12.0\,M_\odot$) with large total cold and
molecular gas reservoirs ($\log{M_{\mathrm{gas}}} \sim 12\,M_\odot$),
in disagreement with the declining trend in cold gas mass in less
(stellar) massive objects. These are a handful of objects with SFRs of
up to $\sim 200\,M_\odot \mathrm{yr}^{-1}$ and drive the increment of
the 50 percentiles in the highest redshift bins. The objects are ultra
luminous infrared galaxies (ULIRGS) with bright 24 $\mu$m intensities
($\sim0.5$ mJy), and are likely to have an AGN component. There are
more ULIRGS in our sample, but it is the combination of high stellar
mass and SFR which makes these galaxies show up so prominently. If
these galaxies indeed have an obscured AGN component, both the stellar
mass and SFR could be overestimated which would be reflected in the cold
and molecular gas content of these objects \citep{caputi2006}.

\begin{figure*}
\includegraphics[width = 1.0\hsize]{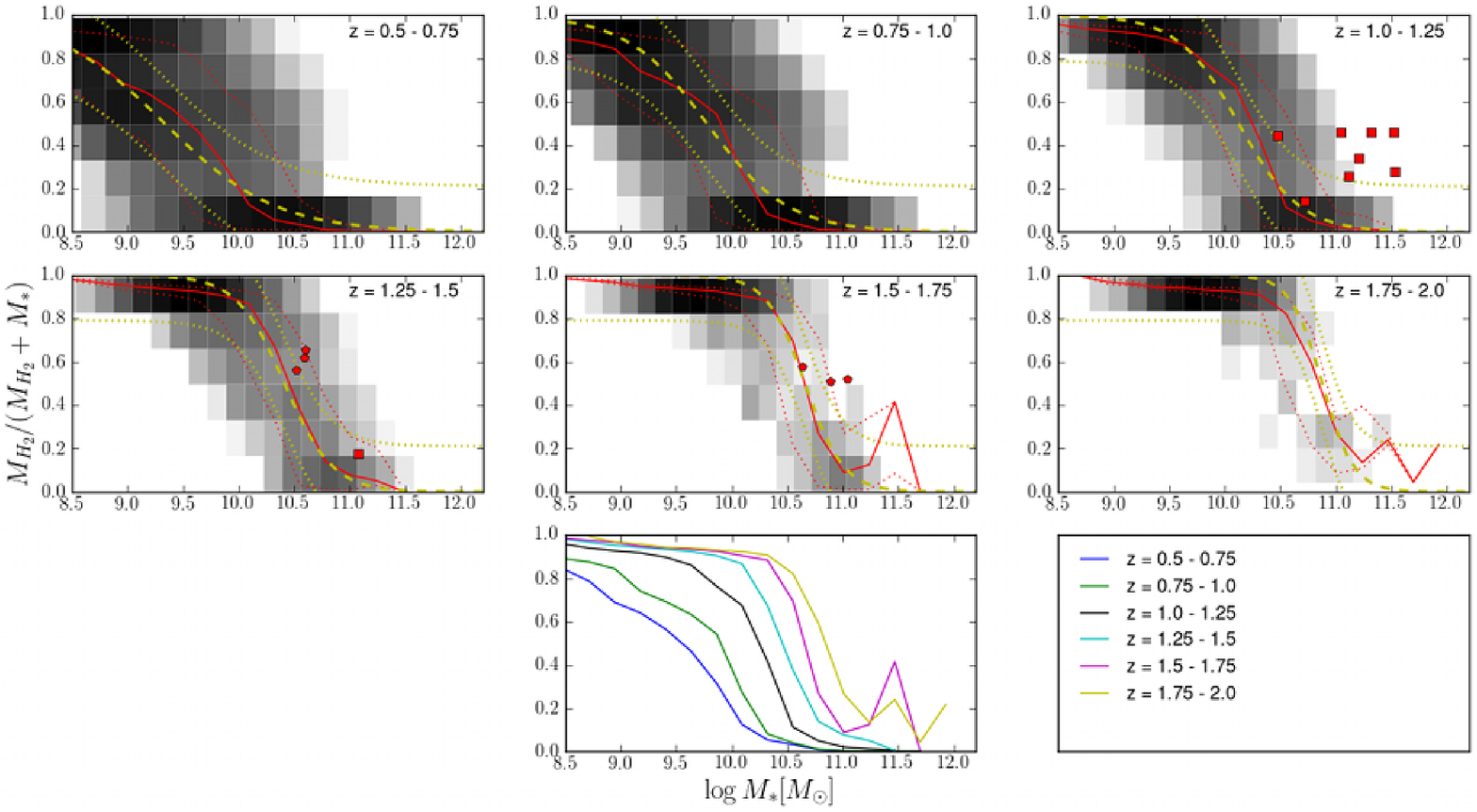}
\caption{$M_{\mathrm{H}_2}/(M_{\mathrm{H}_2} + M_*)$ as a function of
  stellar mass for different redshift bins. The grey shaded area shows
  the log of the number of galaxies in each bin, with the 50, 16 and
  84 percentile curves shown with the red solid and dashed lines. The
  central bottom panel shows the 50 percentile curve for the data in
  each redshift bin. Red squares and pentagons are literature values
  taken from \citet{Tacconi2010} and \citet{Daddi2010},
  respectively. The dashed and dotted yellow lines represents the fit
  and scatter (one sigma) of Equation \ref{eq:H2_frac_fit},
  respectively.\label{fig:H2_gas_frac}}
\end{figure*}

\subsection{Gas fraction}

We present the gas fraction (fraction of all the baryonic mass which
is in cold gas) of our galaxy sample as a function of stellar mass for
different redshift bins in Figure \ref{fig:gas_frac}. We find a clear
trend in gas fraction and stellar mass for all redshift bins, with the
highest gas fractions at low stellar masses. The gas fraction
drops towards higher stellar masses and remains fairly
constant around $0.0$--$0.1$. More prominent, however, is the
evolution of this trend with redshift (see the central panel in the
third row of Figure \ref{fig:gas_frac}). The stellar mass at which the
gas fraction drops increases with increasing redshift, or, in other
words, the gas fraction of galaxies at a given stellar mass increases
with increasing redshift.

Figure \ref{fig:H2_gas_frac} shows the fraction of molecular hydrogen
\h2 with respect to the total stellar and molecular gas mass
($M_{\mathrm{H}_2}/(M_{\mathrm{H}_2} + M_*)$) as a function of stellar
mass. This allows for a direct comparison with high-redshift galaxies
from the literature. We find a trend between
$M_{\mathrm{H}_2}/(M_{\mathrm{H}_2} + M_*)$ and stellar mass similar
to the results of Figure \ref{fig:gas_frac}. It also shows that
$M_{\mathrm{H}_2}/(M_{\mathrm{H}_2} + M_*)$ at a fixed stellar mass
increases with redshift. There is a large spread in
$M_{\mathrm{H}_2}/(M_{\mathrm{H}_2} + M_*)$ at fixed stellar mass,
especially in the lower redshift bins.  As in Figure
\ref{fig:cosmos_H2}, we find good agreement between observations in
the redshift range $1.25<z<1.75$ and our results. We find a deficit
in $M_{\mathrm{H}_2}/(M_{\mathrm{H}_2} + M_*)$  compared to observed objects at
lower redshifts ($1<z<1.25$). As explained in the previous subsection
this can be caused by differences in galaxy population. The peak in cold
gas fraction and $M_{\mathrm{H}_2}/(M_{\mathrm{H}_2} + M_*)$ in the
two highest redshift bins at high stellar masses represent the few
galaxies with high SSFR which are classified as ULIRGS.
\begin{figure*}
\includegraphics[width=\hsize]{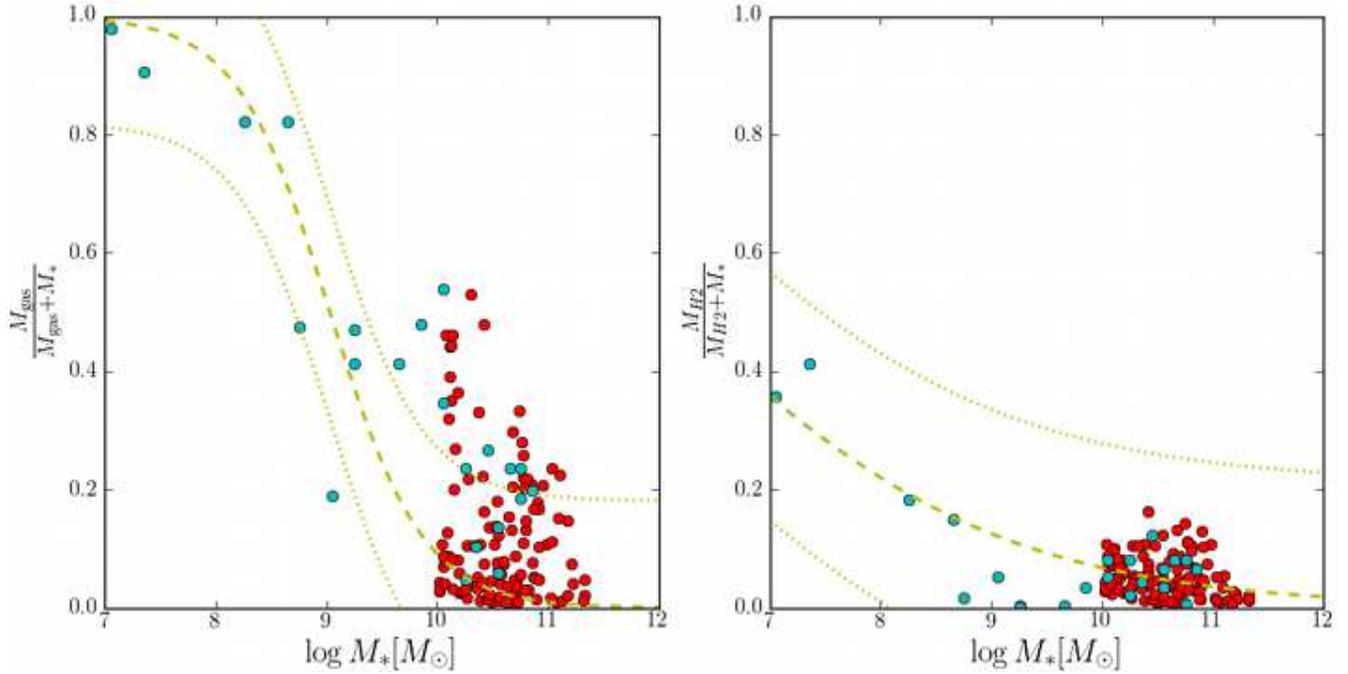}
\caption{Total cold gas fraction (left hand panel) and
  $M_{\mathrm{H}_2}/(M_{\mathrm{H}_2} + M_*)$ (right hand panel)
  vs. stellar mass for local galaxies. Cyan and red dots are from
  \citet{Leroy2008} and \citet{Saintonge2011}, respectively. The
  yellow dashed and dotted lines represent the fit and one sigma
  scatter of Equation \ref{eq:gas_frac_fit} and \ref{eq:H2_frac_fit},
  respectively. These were designed to match our galaxy sample and
  local data. Note that the actual scatter in
  $M_{\mathrm{H}_2}/(M_{\mathrm{H}_2} + M_*)$ for local galaxies is
  much less than the value obtained for Equation
  \ref{eq:H2_frac_fit}.\label{fig:fit_fgas_0.0}}
\end{figure*}

The cold gas fraction in our sampled redshift range can be
characterized by a function of stellar mass and redshift, i.e.,
\begin{equation}
\label{eq:gas_frac_fit}
\frac{M_{\mathrm{gas}}}{M_{\mathrm{gas}} + M_*} = \frac{1}{\exp^{(\log{M_*} - A)/B} + 1},
\end{equation}
where $A = 9.03(1 + \frac{z}{0.35})^{0.11}$ and $B =
0.42(1+z)^{-0.56}$. This fitting formula was designed to match the 50
percentiles in 50 redshifts bins between $0.5<z<1.7$ (where we sampled
a large number of galaxies) and local data from \citet{Leroy2008} and
\citet{Saintonge2011}, respectively. We included local data with the
aim of producing a representation of galaxy gas fraction valid for
both local as distant galaxies. We find an approximately constant
scatter of $\sigma = 0.18$.  This slightly decreases towards the
highest redshifts, but is likely an underestimation due to the low
number of massive objects with large variations in their gas
fractions. Although the function was not designed to match the highest
redshift objects, it does a good job in reproducing their gas
fraction.  The left-hand panel of Fig. \ref{fig:fit_fgas_0.0} shows
the distribution of gas fraction with mass of local galaxies. We
superimpose our fitting function and find that it describes the
observed data fairly well.

Using the same approach we designed a fitting function to
$M_{\mathrm{H}_2}/(M_{\mathrm{H}_2} + M_*)$, which is given by a
similar equation
\begin{equation}
\label{eq:H2_frac_fit}
\frac{M_{\mathrm{H}_2}}{M_{\mathrm{H}_2} + M_*} = \frac{1}{\exp^{(\log{M_*} - A)/B} + 1},
\end{equation}
where $A = 6.15(1 + \frac{z}{0.036})^{0.144}$ and $B =
1.47(1+z)^{-2.23}$.  This formula is a good quantitative
representation of our full sample for local galaxies up to $z=2$, and
has a scatter of $\sigma = 0.21$ for our high-redshift galaxy sample (much less for
the local objects).

Note that at stellar masses above $\log{M_*} > 11.0 \,[M_\odot]$
  the presented fitting formulae do not apply for active
  star forming galaxies, but rather for an 'overall' galaxy population
  which consists of a mixture of star forming and quenched galaxies. We
  advise the reader to use the method presented in section
  \ref{sec:method} in the regime of massive, actively star forming galaxies.

\begin{figure*}
\includegraphics[width = 1.0\hsize]{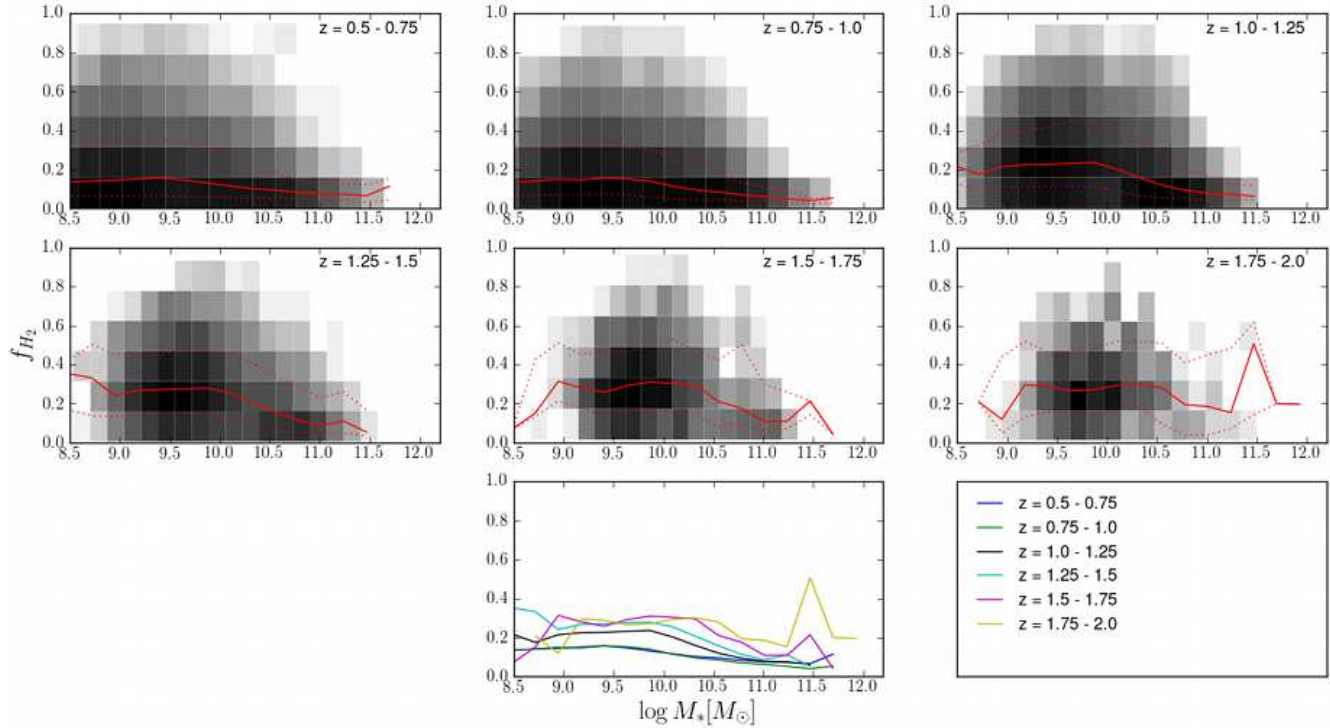}
\caption{The fraction of \h2 in cold gas as a function of stellar mass
  for different redshift bins. The grey shaded area shows the log of
  the number of galaxies in each bin, with the 50, 16 and 84
  percentile curves shown with the red solid and dashed lines. The
  central bottom panel shows the 50 percentile curve for the data in
  each redshift bin.\label{fig:h2_frac_star}}
\end{figure*}

\begin{figure*}
\includegraphics[width = 1.0\hsize]{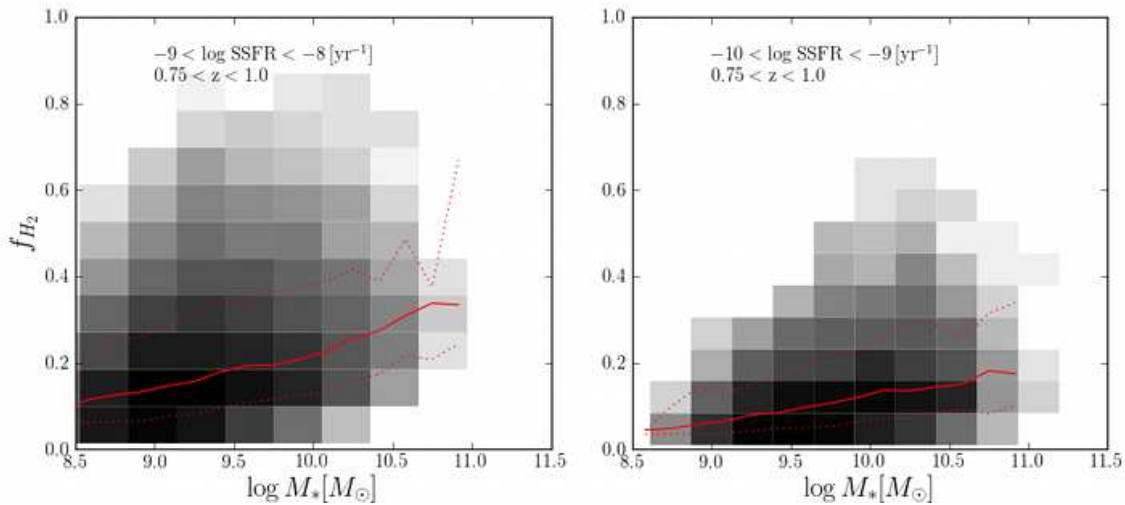}
\caption{The fraction of \h2 in cold gas as a function of stellar mass
  for galaxies with a fixed SSFR. All galaxies have redshifts
  $0.75<\mathrm{z}<1.0$. The grey shaded area shows the log
  of the number of galaxies in each bin, with the 50, 16 and 84
  percentile curves shown with the red solid and dashed lines. Left
  panel: galaxies with SSFR $-9 < \log{\,\mathrm{SSFR}}<-8\,[\mathrm{yr}^{-1}]$. Right panel: galaxies with SSFR $-10 <
  \log{\,\mathrm{SSFR}}<-9\,[\mathrm{yr}^{-1}]$. Note that $f_{H_2}$ decreases with
  decreasing stellar mass. \label{fig:star.vs.fh2_ssfr}}
\end{figure*}

\begin{figure*}
\includegraphics[width = 1.0\hsize]{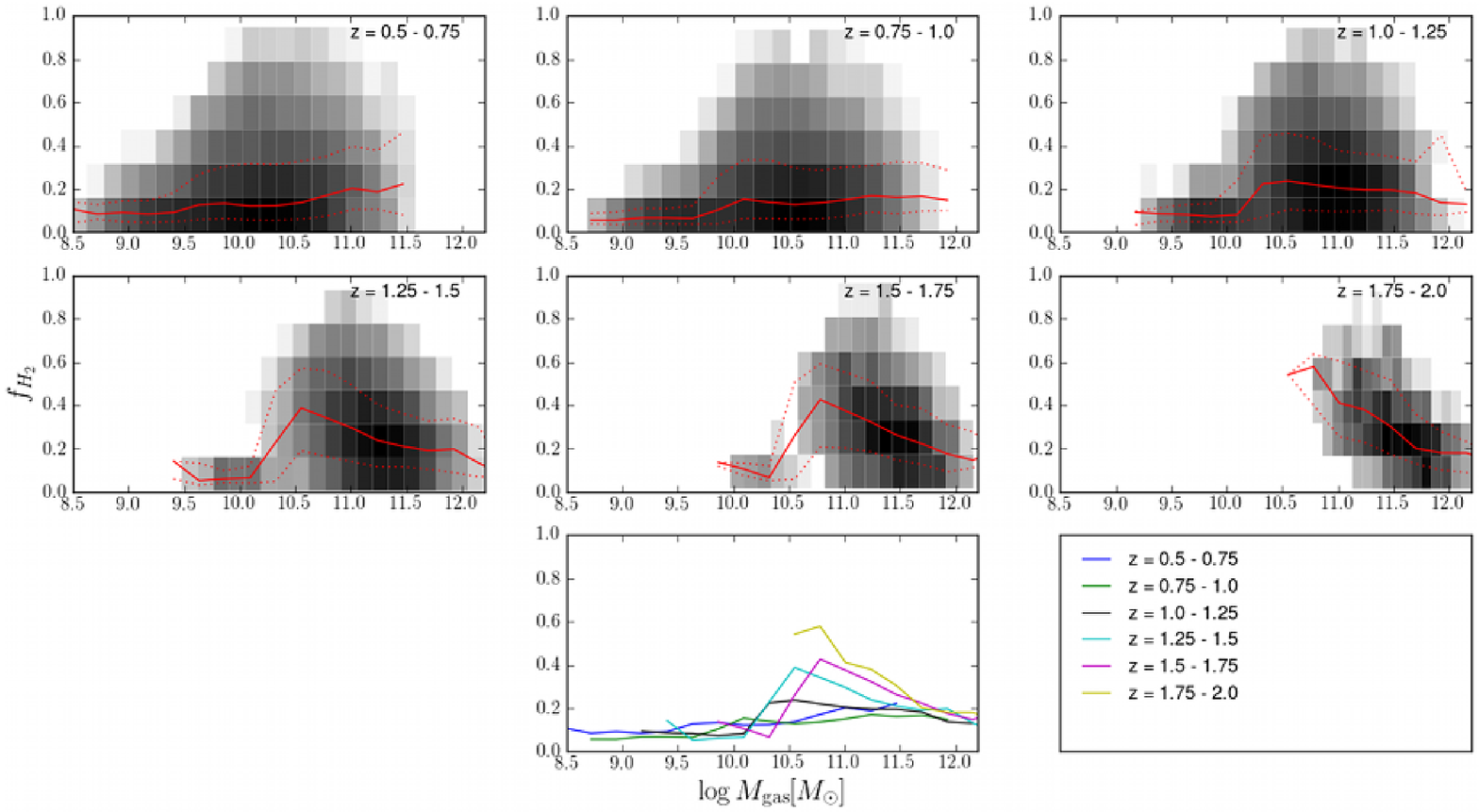}
\caption{The fraction of \h2 in cold gas as a function of total gas
  mass for different redshift bins. The grey shaded area shows the log
  of the number of galaxies in each bin, with the 50, 16 and 84
  percentile curves shown with the red solid and dashed lines. The
  central bottom panel shows the 50 percentile curve for the data in
  each redshift bin. \label{fig:h2_frac_gas}}
\end{figure*}

\subsection{\h2 fraction}

Figure \ref{fig:h2_frac_star} shows the \h2 fraction of the cold gas
(fraction of cold gas which is in a molecular state, $\mathrm{H}_2/(\mathrm{H}_2 +
\mathrm{HI})$) in our galaxy sample as a function of galaxy stellar mass, divided in
different redshift bins. Galaxies have \h2 fractions spanning a range
from nearly zero (hardly any \h2) to almost one (all the gas is in
molecules) at all redshifts. There is nevertheless a (median) trend for
the overall population of the galaxies.

We find the molecular fraction of the gas in most galaxies up to $z=1$
to be around $f_{H_2} = 0.2$, with a minor decrease of the \h2
fraction with increasing stellar mass. The \h2 fraction increases up
to values of $f_{H_2} = 0.3$--$0.4$ for most low-mass objects (below
$\log{M_*/M_\odot}\, \sim 10$) at intermediate redshifts ($z =
1$--$1.5$) and remains at this level at higher redshifts. The
most-massive objects at the highest redshifts ($z = 1.5$--$2.0$) have
molecular gas fractions of $f_{H_2} = 0.2$--$0.3$ and, similar to the
galaxy gas fraction, show a small peak at
$\log{M_*/M_\odot}\sim\,11.5$--$12.0$ (caused by the ULIRGS). In
general, massive galaxies reach low $f_{\mathrm{H}_2}$ values before
less massive galaxies do. This trend is connected to the evolution of
the gas fraction discussed in the previous subsection.

The described trend is reversed compared to the trend observed for
local galaxies \citep[i.e., $f_{H_2}$ decreases with
decreasing stellar mass, e.g.,][]{Leroy2008,Saintonge2011}. This disagreement
is driven by the quenched galaxies (with relatively low SFR and
consequently relative low molecular content) which are dominant towards the massive end of our
galaxy sample. We can decouple the
 mixture of passive and actively star-forming galaxies by considering
 galaxies at fixed SSFR.  We should then expect to see the trend in
 $f_{H_2}$ with stellar mass found in local galaxies, precisely as
 seen in Figure \ref{fig:star.vs.fh2_ssfr}.

The molecular fraction of the cold gas with respect to the total cold
gas in the galaxies for different redshift bins is presented in Figure
\ref{fig:h2_frac_gas}. The figure shows a peak in \h2 fractions at
higher redshift. At the lowest redshifts we see a gentle increase of \h2
fraction with increasing total cold gas mass, whereas at higher
redshift the \h2 fraction peaks at a total cold gas mass of
$\log{M_{\mathrm{gas}}} \sim 10.7$ and decreases at higher masses.

This behavior demonstrates that the apportioning of the cold gas into
atomic and molecular hydrogen is not driven by the cold gas budget
available. This is not surprising as the \citet{Blitz2006} formalism
is based on the cold gas surface density.

\subsection{Transition between gas and stellar dominated phase}
\label{sec:transition}

\begin{figure}
\includegraphics[width = \hsize]{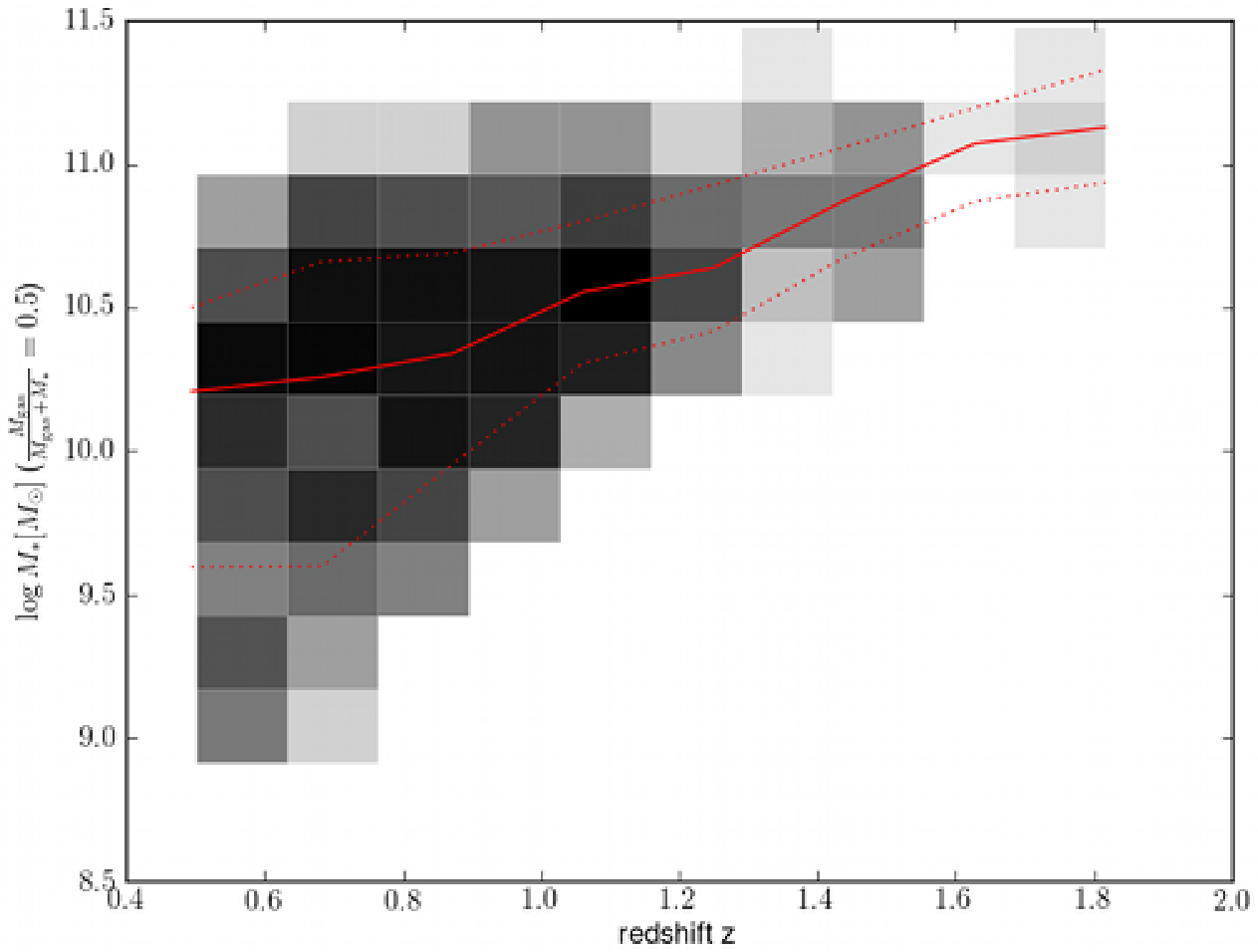}
\caption{Stellar masses of galaxies with $f_{\mathrm{gas}}$ between
  0.47 and 0.53 as a function of redshift. Grey scales show the log of
  the number of galaxies in each bin. Massive galaxies reach the
  transition point from a gas to a stellar dominated phase before less
  massive objects do.\label{fig:gas_0.5_evol}}
\end{figure}

The transition from a gas-dominated phase to a stellar-dominated phase
can be used as a probe of the evolutionary stage of a galaxy. We plot
the stellar masses of galaxies with gas fractions around $\sim$ 0.5
(which marks the transition from gas to stellar dominated) as a
function of redshift in Figure \ref{fig:gas_0.5_evol}. We find that
the typical stellar mass at this transition point decreases from
$\log{M_*(f_{\mathrm{gas}}/M_\odot} = 0.5) = 11.1$ at $z = 2.0$ to
$\log{M_*(f_{\mathrm{gas}}/M_\odot} = 0.5) = 10.1$ at $z = 0.5$. This
decrease of an order of magnitude in stellar mass indicates that
massive galaxies reach the transition from a gas- to a
stellar-dominated phase well before less-massive galaxies.

\section{Discussion}
\label{sec:discussion}

We have shown that one can fairly accurately predict the total cold
gas content of star forming galaxies as a function of their SFR,
stellar mass and size. The cold gas can further be apportioned into
atomic and molecular hydrogen. The key ingredient is a
molecular gas-based star formation law combined with a prescription
for the calculation of cold gas molecular fractions. In order to model
a galaxy, the method assumes an exponential disc, integrated outwards
to five times the stellar disc scale length. We obtain the best
results when increasing the SFE in high cold gas surface density
regions, especially important for high-redshift galaxies (see
Sec. \ref{sec:changing_parameters}). This is in agreement with
previous studies suggesting an increased SFE for dense high-redshift
objects \citep{Daddi2010,Genzel2010}. It is outside the scope of this
work to address the physical origin of this change in SFE.

Our method provides significantly better predictions of galaxy gas
estimates than those derived from the KS law, as can be seen from the
comparison with direct galaxy gas measurements extracted from the
literature. This is due to the break-down of the KS law at low surface
densities, resulting in an underestimation of the cold gas
necessary to sustain low SFR-rates. This break-down leads to
significant changes when integrating over the outer parts of
exponential discs.

Our method is applicable to low- and high-surface-density regions, as
well as low- and high-redshift objects. Furthermore, the empirical
success of the method gives support to our assumption of exponential
profiles for the stellar and gaseous discs. Conversely, the method
developed seems to be a good approach to compute SFRs from gas and
stellar masses.

Despite the success of our method, it fails to properly predict the
molecular and total cold gas content of dwarf galaxies. As stated
previously, \citet{Leroy2008} argue for an unaccounted-for reservoir
of \h2 in these galaxies, possibly due to variations in the $X_{CO}$
factor for dwarfs with respect to bigger spiral galaxies. It would
therefore be worthwhile to address the $X_{CO}$ conversion factor in
dwarfs, in order to better understand the gas physics driving the
partition of cold gas into atomic and molecular hydrogen. This will be
crucial for a proper understanding of star formation at different
stellar masses.

It is worthwhile to remember that we assume that the $X_{CO}$
factor and IMF are universal. An evolving or variable IMF and/or
$X_{CO}$ could potentially lead to significant biases in our
estimates.  A much lower $X_{CO}$ \citep[suggested for starburst
galaxies in e.g.,][]{Genzel2010} would result in much smaller \h2
reservoirs than predicted by our method.  The IMF enters in
converting the observed luminosities to stellar masses, as well as
converting the observational tracers (such as H-$\alpha$, UV or IR
luminosity) to star formation rate. A top-heavy IMF would result in
higher UV, IR or H-$\alpha$ luminosities for a given SFR, or
correspondingly lower SFR estimates for a fixed observed
luminosity. This would result in lower derived gas fractions.

Although the gas masses and molecular fractions presented here are
obtained indirectly, our method allows for the first time the
quantification of the cold gas and molecular gas for a large sample of
galaxies at high redshift. This allows us to study the evolution of
gas content in galaxies and track the relation between
stellar and gas mass. High-redshift star-forming objects appear to be
more gas-rich than lower redshift objects with similar stellar masses
(Fig. \ref{fig:cosmos_gas}). We found that objects at $z \sim 2$ have
a cold gas reservoir approximately thirty times larger than objects at
$z \sim 0.5$ ($\log{M_{\mathrm{gas}}/M_\odot} \sim 11.5$ versus
$\log{M_{\mathrm{gas}}/M_\odot}\sim 10.0$ at a stellar mass of $\log{M_*/M_\odot} =
8.5$) and that the largest gas reservoirs are found in
progressively less-massive galaxies as a function of
time. molecular gas reservoirs follow this behavior closely. These
results indicate that massive galaxies consume or expel their gas earlier and
at a higher rate than less-massive galaxies.

The galaxy gas fractions share the same evolution as total cold gas
contents and decrease rapidly with decreasing redshift
(Fig. \ref{fig:gas_frac}). This trend is in good agreement with the
few direct high redshift molecular gas measures available to date
(Fig. \ref{fig:H2_gas_frac}). More striking is that massive galaxies
reach low gas fractions (less than $\sim 0.1$) much sooner than less-massive
objects.

We find a weak trend between stellar mass and the fraction of gas in
molecular form (see Figure \ref{fig:h2_frac_star}), such that massive
galaxies have somewhat {\em lower} molecular fractions. In addition,
there is a weak redshift trend in the sense that, at a fixed stellar
mass, galaxies have higher molecular fractions at earlier times.

The apparent weak trend between stellar mass and the fraction of gas in
molecular form is driven by the mixture of galaxies on the
main-sequence and already quenched galaxies. We find that the
molecular fraction of cold gas increases with stellar mass when we
decouple this mixture (see Figure \ref{fig:star.vs.fh2_ssfr}). This is
in agreement with observations of local galaxies
\citep{Leroy2008,Saintonge2011}.

We find no clear relationship between the molecular gas fraction and
the total mass of cold gas available (see Figure
\ref{fig:h2_frac_gas}). A large gas reservoir does not necessarily
imply a high molecular fraction.  However, we do find that the
molecular fraction of a galaxy with fixed cold gas mass decreases with
decreasing redshift.
Under the assumption that cold gas is only eligible for star
formation in a molecular state, these results indicate that high-redshift objects with a fixed cold gas content can transform more
cold gas into stars than low redshift objects.

As discussed in Section \ref{sec:changing_parameters} and \ref{sec:sample}, part of our galaxy sample falls off the
  main-sequence and is best described by a Vaucouleurs light profile rather
  than an exponential disc. These objects push to the limits of our
  method (which assumes an exponential disc) and one should interpret these with care.

In a hierarchical galaxy formation scenario, galaxy gas fractions
represent the competition between gas inflow, outflow and
consumption through SF \citep{dave2011}. All the previously
discussed processes hint at a similar general behavior of this
competition with respect to host galaxy redshift and stellar
mass. Massive galaxies consume or expel their gas before
less-massive galaxies do. At a fixed stellar mass galaxies consume
or expel more gas than they attract. This is all in agreement with
\citet{dave2011}, who argue that the cosmic inflow rate of gas
diminishes faster than the consumption rate.  Furthermore, we find
that the molecular fraction of the cold gas decreases with both
stellar mass and time.

We present the transition from a gas to a stellar dominated galaxy in
Figure \ref{fig:gas_0.5_evol}. Galaxies that pass this transition
point have expelled or consumed most of their gas and are likely
(unless they accrete a fresh supply of gas) to proceed towards a
more quiescent phase of SF. We find that more massive galaxies reach a
quiescent gas-poor state before less massive galaxies.

These processes are a reflection of downsizing in star formation,
which can be defined as the declining mass of star-forming galaxies
with decreasing redshift \citep{fontanot2009}.
Our results suggest that the slowly shrinking gas reservoirs, combined
with lower molecular gas fractions, may drive the downsizing observed
in star formation. It will be interesting to test this picture with
future `direct' observations of gas in galaxies at high redshift.

\section{Summary and Conclusion}
\label{sec:summary}

We developed a method to indirectly measure the total cold gas and
molecular gas content of galaxies. We applied this method to a sample
of galaxies from the \cosmos survey covering a redshift range of $0.5
\leq z \leq 2.0$. Since we obtained galaxy gas masses from their SFR,
our results are effectively another way of representing SF and might
seem redundant with other previous studies in the literature. It is
therefore worthwhile to briefly summarize what we actually learned
from this approach that we could not have concluded from the SFR
alone.  Our main results are:

\begin{itemize}
\item The gas content of a galaxy can be described using a combination
  of a molecular SF-law and a prescription to calculate the molecular
  fraction of cold gas, all under the assumption of a radial
  exponential profile for the gas and stars. Best results are obtained
  using a variable SFE. Conversely, this method can act as a scaling
  law relating the SFR surface density to the cold gas surface density
  of a galaxy. The method seems to be applicable to a large range of
  galaxy gas masses at both low and high redshift.

\item The method presented allows us not only to predict the cold gas
  content but also to predict the molecular gas content of galaxies,
  and it is based on observed galaxy properties. These estimates can
  help us to interpret direct observations of molecular gas in
  galaxies at high redshift relative to the overall galaxy population. 

\item We find a strong trend between the gas content of a galaxy and
  its stellar mass with time. For a given stellar mass, the gas
  fraction of the galaxy decreases with decreasing redshift.  On
  average, massive galaxies consume or expel their gas reservoir much
  earlier in the history of the Universe than less massive galaxies.
  This trend is not only observed in the cold gas content and gas
  fraction of the galaxies; it is also reflected in the molecular
  fraction of the cold gas, which gets smaller with time for a given
  stellar mass. This, in combination with low gas fractions, shows
  that the physics which suppress the formation of stars in massive
  galaxies with time is at least two-fold: galaxies run out of gas
  \emph{and} molecules, but not necessarily at the same rate.

\item  The stellar mass at which galaxies become stellar-dominated
  decreases with time. This indicates that massive galaxies reach a
  gas-poor state before less massive objects.

\item These results point towards a common trend: the more massive a
  galaxy, the sooner it consumes or expels its gas content. This trend
  is another manifestation of downsizing.

\end{itemize}

Although still indirect, these results point out that measuring the
gas content of galaxies is of great use for developing a broader
picture and better quantification of galaxy evolution. The method and
results presented in this work are a first step in this direction and
can help for the design of upcoming surveys. Furthermore, they provide
useful constraints for cosmological galaxy formation models,
particularly for the development of new models which include detailed
tracking of multi-phase gas.


\section*{Acknowledgments}
We thank Fran\c{c}oise Combes and Andrew Baker for stimulating
discussions, and Marco Spaans for a reading of the manuscript. GP
acknowledges NOVA (Nederlandse Onderzoekschool Voor Astronomie) for
funding. KIC acknowledges the Leverhulme Trust for funding through the
award of an Early Career Fellowship while this work was being done.

\bibliographystyle{mn2e_fix}
\bibliography{references}

\begin{thebibliography}{54}
\expandafter\ifx\csname natexlab\endcsname\relax\def\natexlab#1{#1}\fi

\bibitem[{{Bigiel} {et~al}\mbox{.}(2008){Bigiel}, {Leroy}, {Walter}, {Brinks},
  {de Blok}, {Madore}, \& {Thornley}}]{Bigiel2008}
{Bigiel} F., {Leroy} A., {Walter} F., {Brinks} E., {de Blok} W.~J.~G., {Madore}
  B., {Thornley} M.~D., 2008, \aj, 136, 2846

\bibitem[{{Bigiel} {et~al}\mbox{.}(2011){Bigiel}, {Leroy}, {Walter}, {Brinks},
  {de Blok}, {Kramer}, {Rix}, {Schruba}, {Schuster}, {Usero}, \&
  {Wiesemeyer}}]{Bigiel2011}
{Bigiel} F. {et~al.}, 2011, \apjl, 730, L13+

\bibitem[{{Blitz} \& {Rosolowsky}(2004)}]{blitz2004}
{Blitz} L., {Rosolowsky} E., 2004, \apjl, 612, L29

\bibitem[{{Blitz} \& {Rosolowsky}(2006)}]{Blitz2006}
---, 2006, \apj, 650, 933

\bibitem[{{Bolatto} {et~al}\mbox{.}(2008){Bolatto}, {Leroy}, {Rosolowsky},
  {Walter}, \& {Blitz}}]{Bolatto2008}
{Bolatto} A.~D., {Leroy} A.~K., {Rosolowsky} E., {Walter} F., {Blitz} L., 2008,
  \apj, 686, 948

\bibitem[{{Bottema}(1993)}]{Bottema1993}
{Bottema} R., 1993, \aap, 275, 16

\bibitem[{{Broeils} \& {Rhee}(1997)}]{broeils1997}
{Broeils} A.~H., {Rhee} M.-H., 1997, \aap, 324, 877

\bibitem[{{Bruzual}(2007)}]{bruzual2007}
{Bruzual} G., 2007, in Astronomical Society of the Pacific Conference Series,
  Vol. 374, From Stars to Galaxies: Building the Pieces to Build Up the
  Universe, {A.~Vallenari, R.~Tantalo, L.~Portinari, \& A.~Moretti}, ed., pp.
  303--+

\bibitem[{{Caputi} {et~al}\mbox{.}(2006){Caputi}, {Dole}, {Lagache}, {McLure},
  {Puget}, {Rieke}, {Dunlop}, {Le Floc'h}, {Papovich}, \&
  {P{\'e}rez-Gonz{\'a}lez}}]{caputi2006}
{Caputi} K.~I. {et~al.}, 2006, \apj, 637, 727

\bibitem[{{Chabrier}(2003)}]{chabrier2003}
{Chabrier} G., 2003, \pasp, 115, 763

\bibitem[{{Cimatti}, {Daddi} \& {Renzini}(2006){Cimatti}, {Daddi}, \&
  {Renzini}}]{Cimatti2006}
{Cimatti} A., {Daddi} E., {Renzini} A., 2006, \aap, 453, L29

\bibitem[{{Cowie} {et~al}\mbox{.}(1996){Cowie}, {Songaila}, {Hu}, \&
  {Cohen}}]{Cowie1996}
{Cowie} L.~L., {Songaila} A., {Hu} E.~M., {Cohen} J.~G., 1996, \aj, 112, 839

\bibitem[{{Daddi} {et~al}\mbox{.}(2010){Daddi}, {Bournaud}, {Walter},
  {Dannerbauer}, {Carilli}, {Dickinson}, {Elbaz}, {Morrison}, {Riechers},
  {Onodera}, {Salmi}, {Krips}, \& {Stern}}]{Daddi2010}
{Daddi} E. {et~al.}, 2010, \apj, 713, 686

\bibitem[{{Dav{\'e}}, {Finlator} \& {Oppenheimer}(2011){Dav{\'e}}, {Finlator},
  \& {Oppenheimer}}]{dave2011}
{Dav{\'e}} R., {Finlator} K., {Oppenheimer} B.~D., 2011, \mnras, 416, 1354

\bibitem[{{Drory} {et~al}\mbox{.}(2004){Drory}, {Bender}, {Feulner}, {Hopp},
  {Maraston}, {Snigula}, \& {Hill}}]{Drory2004}
{Drory} N., {Bender} R., {Feulner} G., {Hopp} U., {Maraston} C., {Snigula} J.,
  {Hill} G.~J., 2004, \apj, 608, 742

\bibitem[{{Drory} {et~al}\mbox{.}(2005){Drory}, {Salvato}, {Gabasch}, {Bender},
  {Hopp}, {Feulner}, \& {Pannella}}]{Drory2005}
{Drory} N., {Salvato} M., {Gabasch} A., {Bender} R., {Hopp} U., {Feulner} G.,
  {Pannella} M., 2005, \apjl, 619, L131

\bibitem[{{Elmegreen}(1989)}]{Elmegreen1989}
{Elmegreen} B.~G., 1989, \apj, 338, 178

\bibitem[{{Elmegreen}(1993)}]{Elmegreen1993}
---, 1993, \apj, 411, 170

\bibitem[{{Erb} {et~al}\mbox{.}(2006){Erb}, {Steidel}, {Shapley}, {Pettini},
  {Reddy}, \& {Adelberger}}]{Erb2006}
{Erb} D.~K., {Steidel} C.~C., {Shapley} A.~E., {Pettini} M., {Reddy} N.~A.,
  {Adelberger} K.~L., 2006, \apj, 646, 107

\bibitem[{{Faber}, {Worthey} \& {Gonzales}(1992){Faber}, {Worthey}, \&
  {Gonzales}}]{Faber1992}
{Faber} S.~M., {Worthey} G., {Gonzales} J.~J., 1992, in IAU Symposium, Vol.
  149, The Stellar Populations of Galaxies, {B.~Barbuy \& A.~Renzini}, ed., pp.
  255--+

\bibitem[{{Fontanot} {et~al}\mbox{.}(2009){Fontanot}, {De Lucia}, {Monaco},
  {Somerville}, \& {Santini}}]{fontanot2009}
{Fontanot} F., {De Lucia} G., {Monaco} P., {Somerville} R.~S., {Santini} P.,
  2009, \mnras, 397, 1776

\bibitem[{{Fu} {et~al}\mbox{.}(2010){Fu}, {Guo}, {Kauffmann}, \&
  {Krumholz}}]{Fu2010}
{Fu} J., {Guo} Q., {Kauffmann} G., {Krumholz} M.~R., 2010, \mnras, 409, 515

\bibitem[{{Genzel} {et~al}\mbox{.}(2010){Genzel}, {Tacconi}, {Gracia-Carpio},
  {Sternberg}, {Cooper}, {Shapiro}, {Bolatto}, {Bouch{\'e}}, {Bournaud},
  {Burkert}, {Combes}, {Comerford}, {Cox}, {Davis}, {Schreiber},
  {Garcia-Burillo}, {Lutz}, {Naab}, {Neri}, {Omont}, {Shapley}, \&
  {Weiner}}]{Genzel2010}
{Genzel} R. {et~al.}, 2010, \mnras, 407, 2091

\bibitem[{{Gnedin} \& {Kravtsov}(2011)}]{Gnedin2011}
{Gnedin} N.~Y., {Kravtsov} A.~V., 2011, \apj, 728, 88

\bibitem[{{Hopkins}(2004)}]{Hopkins2004}
{Hopkins} A.~M., 2004, \apj, 615, 209

\bibitem[{{Hopkins} \& {Beacom}(2006)}]{Hopkins2006}
{Hopkins} A.~M., {Beacom} J.~F., 2006, \apj, 651, 142

\bibitem[{{Ilbert} {et~al}\mbox{.}(2009){Ilbert}, {Capak}, {Salvato}, {Aussel},
  {McCracken}, {Sanders}, {Scoville}, {Kartaltepe}, {Arnouts}, {Le Floc'h},
  {Mobasher}, {Taniguchi}, {Lamareille}, {Leauthaud}, {Sasaki}, {Thompson},
  {Zamojski}, {Zamorani}, {Bardelli}, {Bolzonella}, {Bongiorno}, {Brusa},
  {Caputi}, {Carollo}, {Contini}, {Cook}, {Coppa}, {Cucciati}, {de la Torre},
  {de Ravel}, {Franzetti}, {Garilli}, {Hasinger}, {Iovino}, {Kampczyk},
  {Kneib}, {Knobel}, {Kovac}, {Le Borgne}, {Le Brun}, {F{\`e}vre}, {Lilly},
  {Looper}, {Maier}, {Mainieri}, {Mellier}, {Mignoli}, {Murayama}, {Pell{\`o}},
  {Peng}, {P{\'e}rez-Montero}, {Renzini}, {Ricciardelli}, {Schiminovich},
  {Scodeggio}, {Shioya}, {Silverman}, {Surace}, {Tanaka}, {Tasca}, {Tresse},
  {Vergani}, \& {Zucca}}]{Ilbert2009}
{Ilbert} O. {et~al.}, 2009, \apj, 690, 1236

\bibitem[{{Kennicutt}(1998{\natexlab{a}})}]{Kennicutt1998SFR}
{Kennicutt}, Jr. R.~C., 1998{\natexlab{a}}, \araa, 36, 189

\bibitem[{{Kennicutt}(1998{\natexlab{b}})}]{Kennicutt1998law}
---, 1998{\natexlab{b}}, \apj, 498, 541

\bibitem[{{Krumholz} \& {Dekel}(2011)}]{Krumholz2011}
{Krumholz} M.~R., {Dekel} A., 2011, ArXiv e-prints

\bibitem[{{Kuhlen} {et~al}\mbox{.}(2011){Kuhlen}, {Krumholz}, {Madau}, {Smith},
  \& {Wise}}]{Kuhlen2011}
{Kuhlen} M., {Krumholz} M., {Madau} P., {Smith} B., {Wise} J., 2011, ArXiv
  e-prints

\bibitem[{{Lagos} {et~al}\mbox{.}(2011{\natexlab{a}}){Lagos}, {Baugh}, {Lacey},
  {Benson}, {Kim}, \& {Power}}]{Lagos2011cosmic_evol}
{Lagos} C.~D.~P., {Baugh} C.~M., {Lacey} C.~G., {Benson} A.~J., {Kim} H.-S.,
  {Power} C., 2011{\natexlab{a}}, \mnras, 1776

\bibitem[{{Lagos} {et~al}\mbox{.}(2011{\natexlab{b}}){Lagos}, {Lacey}, {Baugh},
  {Bower}, \& {Benson}}]{Lagos2011sflaw}
{Lagos} C.~D.~P., {Lacey} C.~G., {Baugh} C.~M., {Bower} R.~G., {Benson} A.~J.,
  2011{\natexlab{b}}, \mnras, 416, 1566

\bibitem[{{Leroy} {et~al}\mbox{.}(2008){Leroy}, {Walter}, {Brinks}, {Bigiel},
  {de Blok}, {Madore}, \& {Thornley}}]{Leroy2008}
{Leroy} A.~K., {Walter} F., {Brinks} E., {Bigiel} F., {de Blok} W.~J.~G.,
  {Madore} B., {Thornley} M.~D., 2008, \aj, 136, 2782

\bibitem[{{Madau} {et~al}\mbox{.}(1996){Madau}, {Ferguson}, {Dickinson},
  {Giavalisco}, {Steidel}, \& {Fruchter}}]{Madau1996}
{Madau} P., {Ferguson} H.~C., {Dickinson} M.~E., {Giavalisco} M., {Steidel}
  C.~C., {Fruchter} A., 1996, \mnras, 283, 1388

\bibitem[{{Mannucci} {et~al}\mbox{.}(2009){Mannucci}, {Cresci}, {Maiolino},
  {Marconi}, {Pastorini}, {Pozzetti}, {Gnerucci}, {Risaliti}, {Schneider},
  {Lehnert}, \& {Salvati}}]{Mannucci2009}
{Mannucci} F. {et~al.}, 2009, \mnras, 398, 1915

\bibitem[{{McKee} \& {Ostriker}(2007)}]{McKee2007}
{McKee} C.~F., {Ostriker} E.~C., 2007, \araa, 45, 565

\bibitem[{{Obreschkow} \& {Rawlings}(2009)}]{Obreschkow2009}
{Obreschkow} D., {Rawlings} S., 2009, \mnras, 394, 1857

\bibitem[{{Saintonge} {et~al}\mbox{.}(2011){Saintonge}, {Kauffmann}, {Kramer},
  {Tacconi}, {Buchbender}, {Catinella}, {Fabello}, {Graci{\'a}-Carpio}, {Wang},
  {Cortese}, {Fu}, {Genzel}, {Giovanelli}, {Guo}, {Haynes}, {Heckman},
  {Krumholz}, {Lemonias}, {Li}, {Moran}, {Rodriguez-Fernandez}, {Schiminovich},
  {Schuster}, \& {Sievers}}]{Saintonge2011}
{Saintonge} A. {et~al.}, 2011, \mnras, 415, 32

\bibitem[{{Salpeter}(1955)}]{salpeter1955}
{Salpeter} E.~E., 1955, \apj, 121, 161

\bibitem[{{Scarlata} {et~al}\mbox{.}(2007){Scarlata}, {Carollo}, {Lilly},
  {Sargent}, {Feldmann}, {Kampczyk}, {Porciani}, {Koekemoer}, {Scoville},
  {Kneib}, {Leauthaud}, {Massey}, {Rhodes}, {Tasca}, {Capak}, {Maier},
  {McCracken}, {Mobasher}, {Renzini}, {Taniguchi}, {Thompson}, {Sheth},
  {Ajiki}, {Aussel}, {Murayama}, {Sanders}, {Sasaki}, {Shioya}, \&
  {Takahashi}}]{Scarlata2007}
{Scarlata} C. {et~al.}, 2007, \apjs, 172, 406

\bibitem[{{Schmidt}(1959)}]{schmidt1959}
{Schmidt} M., 1959, \apj, 129, 243

\bibitem[{{Schruba} {et~al}\mbox{.}(2011){Schruba}, {Leroy}, {Walter},
  {Bigiel}, {Brinks}, {de Blok}, {Dumas}, {Kramer}, {Rosolowsky}, {Sandstrom},
  {Schuster}, {Usero}, {Weiss}, \& {Wiesemeyer}}]{Schruba2011}
{Schruba} A. {et~al.}, 2011, \aj, 142, 37

\bibitem[{{Scoville} {et~al}\mbox{.}(2007){Scoville}, {Aussel}, {Brusa},
  {Capak}, {Carollo}, {Elvis}, {Giavalisco}, {Guzzo}, {Hasinger}, {Impey},
  {Kneib}, {LeFevre}, {Lilly}, {Mobasher}, {Renzini}, {Rich}, {Sanders},
  {Schinnerer}, {Schminovich}, {Shopbell}, {Taniguchi}, \&
  {Tyson}}]{scoville2007}
{Scoville} N. {et~al.}, 2007, \apjs, 172, 1

\bibitem[{{Shostak} \& {van der Kruit}(1984)}]{shostak1984}
{Shostak} G.~S., {van der Kruit} P.~C., 1984, \aap, 132, 20

\bibitem[{{Solomon} {et~al}\mbox{.}(1987){Solomon}, {Rivolo}, {Barrett}, \&
  {Yahil}}]{Solomon1987}
{Solomon} P.~M., {Rivolo} A.~R., {Barrett} J., {Yahil} A., 1987, \apj, 319, 730

\bibitem[{{Spaans} \& {Norman}(1997)}]{spaans1997}
{Spaans} M., {Norman} C.~A., 1997, \apj, 483, 87

\bibitem[{{Tacconi} {et~al}\mbox{.}(2010){Tacconi}, {Genzel}, {Neri}, {Cox},
  {Cooper}, {Shapiro}, {Bolatto}, {Bouch{\'e}}, {Bournaud}, {Burkert},
  {Combes}, {Comerford}, {Davis}, {Schreiber}, {Garcia-Burillo},
  {Gracia-Carpio}, {Lutz}, {Naab}, {Omont}, {Shapley}, {Sternberg}, \&
  {Weiner}}]{Tacconi2010}
{Tacconi} L.~J. {et~al.}, 2010, \nat, 463, 781

\bibitem[{{Tasca} {et~al}\mbox{.}(2009){Tasca}, {Kneib}, {Iovino}, {Le
  F{\`e}vre}, {Kova{\v c}}, {Bolzonella}, {Lilly}, {Abraham}, {Cassata},
  {Cucciati}, {Guzzo}, {Tresse}, {Zamorani}, {Capak}, {Garilli}, {Scodeggio},
  {Sheth}, {Zucca}, {Carollo}, {Contini}, {Mainieri}, {Renzini}, {Bardelli},
  {Bongiorno}, {Caputi}, {Coppa}, {de La Torre}, {de Ravel}, {Franzetti},
  {Kampczyk}, {Knobel}, {Koekemoer}, {Lamareille}, {Le Borgne}, {Le Brun},
  {Maier}, {Mignoli}, {Pello}, {Peng}, {Perez Montero}, {Ricciardelli},
  {Silverman}, {Vergani}, {Tanaka}, {Abbas}, {Bottini}, {Cappi}, {Cimatti},
  {Ilbert}, {Leauthaud}, {Maccagni}, {Marinoni}, {McCracken}, {Memeo},
  {Meneux}, {Oesch}, {Porciani}, {Pozzetti}, {Scaramella}, \&
  {Scarlata}}]{Tasca2009}
{Tasca} L.~A.~M. {et~al.}, 2009, \aap, 503, 379

\bibitem[{{Trager}, {Faber} \& {Dressler}(2008){Trager}, {Faber}, \&
  {Dressler}}]{Trager2008}
{Trager} S.~C., {Faber} S.~M., {Dressler} A., 2008, \mnras, 386, 715

\bibitem[{{Trager} {et~al}\mbox{.}(2000){Trager}, {Faber}, {Worthey}, \&
  {Gonz{\'a}lez}}]{Trager2000}
{Trager} S.~C., {Faber} S.~M., {Worthey} G., {Gonz{\'a}lez} J.~J., 2000, \aj,
  120, 165

\bibitem[{{Wong} \& {Blitz}(2002)}]{wong2002}
{Wong} T., {Blitz} L., 2002, \apj, 569, 157

\bibitem[{{Worthey}, {Faber} \& {Gonzalez}(1992){Worthey}, {Faber}, \&
  {Gonzalez}}]{Worthey1992}
{Worthey} G., {Faber} S.~M., {Gonzalez} J.~J., 1992, \apj, 398, 69

\bibitem[{{Wuyts} {et~al}\mbox{.}(2011){Wuyts}, {F{\"o}rster Schreiber}, {van
  der Wel}, {Magnelli}, {Guo}, {Genzel}, {Lutz}, {Aussel}, {Barro}, {Berta},
  {Cava}, {Graci{\'a}-Carpio}, {Hathi}, {Huang}, {Kocevski}, {Koekemoer},
  {Lee}, {Le Floc'h}, {McGrath}, {Nordon}, {Popesso}, {Pozzi}, {Riguccini},
  {Rodighiero}, {Saintonge}, \& {Tacconi}}]{wuyts2011}
{Wuyts} S. {et~al.}, 2011, \apj, 742, 96

\end{thebibliography}

\end{document}